\documentclass[11pt, a4paper]{article}
\usepackage[english]{babel}
\usepackage[utf8]{inputenc}
\usepackage{jheppub}

\usepackage{amsmath}
\usepackage{amsfonts}
\usepackage{amssymb}
\usepackage{float}
\usepackage{enumerate}
\usepackage{slashed}

\usepackage{graphicx}
\usepackage[dvipsnames]{xcolor}
\usepackage{mathtools}
\usepackage{simplewick}
\usepackage{hyperref}
\usepackage{multirow}
\usepackage{enumitem}
\usepackage{physics}
\usepackage[compat=1.1.0]{tikz-feynman}
\usepackage{subfigure}
\usepackage{booktabs}
\usepackage{xcolor}
\usepackage{tikz}

\usetikzlibrary{decorations.pathmorphing}
\usepackage[export]{adjustbox}

\usepackage{latexsym}
\usepackage{mathrsfs}
\usepackage{amsthm}
\usepackage{amstext}

\usepackage{graphicx}
\usepackage{slashed}

\makeatletter
\@addtoreset{subfigure}{row}
\makeatother

\usepackage[rightcaption]{sidecap}
\usepackage{caption}

\usepackage{mathdots}
\usepackage{yhmath}
\usepackage{cancel}
\usepackage{color}
\usepackage{siunitx}
\usepackage{array}

\usepackage{tabularx}
\usepackage{extarrows}
\usepackage{booktabs}
\usetikzlibrary{fadings}
\usetikzlibrary{patterns}
\usetikzlibrary{shadows.blur}
\usetikzlibrary{shapes}

\allowdisplaybreaks[1]

\definecolor{ccqqqq}{rgb}{1,0.5,0}
\definecolor{uuuuuu}{rgb}{0.26666666666666666,0.26666666666666666,0.26666666666666666}
\definecolor{qqwwzz}{rgb}{0,0.3,0.9}

\def\p{\partial}

\newcommand{\beq}{\begin{equation}}
\newcommand{\eeq}{\end{equation}}
\newcommand{\bea}{\begin{eqnarray}}
\newcommand{\eea}{\end{eqnarray}}

\def\le{\left(}
\def\ri{\right)}

\numberwithin{equation}{section}

\usepackage{amsthm}

\newtheorem{thm}{Theorem}[section] 
\newtheorem{srule}[thm]{Selection rule}

\def\le{\left(}
\def\ri{\right)}
\newcommand{\eg}{{\it e.g.,}\ }
\newcommand{\ie}{{\it i.e.,}\ }
\newcommand{\SU}{\text{SU}}

\newcommand{\spa}{\ , \ \ }

\usepackage{makecell}

\title{
Finite scalar field theory with SU(1,1) spacetime symmetry 
from near-BPS limits of $\mathcal{N}=4$ SYM
}
\author{Roberto Auzzi$^{1, \, 2}$,  Stefano Baiguera$^{1, \, 2}$, Lihan Guo$^{1, \, 2, \, 3}$, Troels Harmark$^{4, \, 5}$, \\ Giuseppe Nardelli$^{2, \, 6}$}

\affiliation{$^1$INFN Sezione di Perugia, Via A. Pascoli, 06123 Perugia, Italy}
\affiliation{$^2$Dipartimento di Matematica e Fisica, Universit\`{a} Cattolica del Sacro Cuore, \\ Via della Garzetta 48, 25133 Brescia, Italy}
\affiliation{$^3$Institute for Theoretical Physics, KU Leuven, \\
Celestijnenlaan 200D, B-3001 Leuven, Belgium}
\affiliation{$^4$Center of Gravity, Niels Bohr Institute, University of Copenhagen, \\
Blegdamsvej 17, DK-2100 Copenhagen Ø, Denmark}
\affiliation{$^5$Nordita, KTH Royal Institute of Technology and Stockholm University, \\
Hannes Alfvéns väg 12, SE-106 91 Stockholm, Sweden}
\affiliation{$^6$ TIFPA - INFN, c/o Dipartimento di Fisica, Università di Trento, \\ 38123 Povo (TN), Italy }

\emailAdd{roberto.auzzi@unicatt.it}
\emailAdd{stefano.baiguera@pg.infn.it}
\emailAdd{lihan.guo@unicatt.it}
\emailAdd{harmark@nbi.ku.dk}
\emailAdd{giuseppe.nardelli@unicatt.it}

\abstract{In this work, we consider an interacting and matrix-valued scalar quantum field theory that emerges from a near-BPS decoupling limit of $\mathcal{N}=4$ super Yang-Mills.
The theory is non-Lorentzian with SU(1,1) spacetime symmetry and admits a (semi-)local action formulation. 
The interaction can be viewed as arising from a non-abelian gauge field without propagating degrees of freedom. 
The proposed field theory action has previously been considered as classically equivalent to SU(1,1) Spin Matrix theory. 
In this work, we examine this equivalence at the quantum level.
We show that the classical action is off-shell invariant under the SU(1,1) symmetry group.
We then analyze the renormalization properties, showing the theory is finite at all orders in perturbation theory.
This provides a rare example of a non-supersymmetric and non-Lorentzian quantum field theory where a non-renormalization theorem holds.
}

\begin{document}

\maketitle

\setcounter{tocdepth}{1}

\section{Introduction}

The holographic principle, relating gravity in a bulk geometry to a quantum theory on its boundary, led to major achievements in theoretical physics in the last decades, primarily within the well-established framework of the Anti-de Sitter/Conformal Field Theory (AdS/CFT) correspondence~\cite{Maldacena:1997re}.
Despite substantial progress, it remains desirable to find a tractable holographic quantum field theory (QFT) defined in a regime where non-perturbative effects on the string theory side -- such as D-branes and black holes -- are accessible.
Such a controllable framework can be attained by considering certain decoupling limits of $\mathcal{N}=4$ super Yang-Mills (SYM) theory with gauge group $\mathrm{SU}(N)$, such that only a subset of the degrees of freedom in the original theory are retained~\cite{Harmark:2007px,Harmark:2014mpa}.
The models obtained in this way are referred to as \textit{Spin Matrix Theories} (SMTs), since the fields belong to the adjoint representation of $\mathrm{SU}(N)$ and they transform under a spin group, obtained as a subset of the symmetry group $\mathrm{PSU}(2,2|4)$.
SMTs are valuable models that interconnect fundamental physics -- including the study of Hagedorn phase transitions, giant gravitons, non-relativistic string theory and more~\cite{Kinney:2005ej,Harmark:2007px,Harmark:2014mpa,Harmark:2016cjq,Harmark:2017rpg,Harmark:2018cdl,Harmark:2019upf,Harmark:2019zkn,Baiguera:2020jgy,Harmark:2020vll,Baiguera:2020mgk,Baiguera:2021hky,Baiguera:2022pll,Oling:2022fft,Baiguera:2023fus,Bidussi:2023rfs,Demulder:2023bux,Blair:2023noj,Blair:2024aqz,Baiguera:2024vlj} -- with condensed matter applications, such as the Heisenberg XXX$_{1/2}$ spin chain and the Landau-Lifshitz model~\cite{Harmark:2006ta,Harmark:2006ie,Harmark:2008gm,Kruczenski:2003gt,Klose:2006dd}.

In this paper, we focus on the so-called bosonic SU(1,1) SMT, whose dynamical field content is fully characterized by a single complex scalar field $\Phi$.
We will provide a semi-local field theory formulation and study its renormalization properties at quantum level.
We will show that this theory provides a very exceptional case of non-Lorentzian and non-supersymmetric QFT that is finite at all orders in perturbation theory.

\vskip 5mm

The most studied example of the AdS/CFT correspondence relates four-dimensional $\mathcal{N}=4$ SYM theory to type IIB string theory on $\mathrm{AdS}_5 \times S^5$.
Various research directions have been pursued to investigate this case; for instance, by exploiting the power of integrability in the planar limit $N=\infty$~\cite{Minahan:2002ve,Beisert:2003jj,Beisert:2003tq,Beisert:2003ys,Beisert:2004ry,Kazakov:2004qf,Beisert:2007sk,Minahan:2010js,Beisert:2010jr}, or by employing exact techniques such as supersymmetric localization, \eg see~\cite{Pestun:2007rz,Benini:2015eyy,Pestun:2016zxk,Hosseini:2017mds,Cabo-Bizet:2018ehj,Choi:2018hmj,Benini:2018ywd,Zaffaroni:2019dhb,Goldstein:2020yvj}. 
SMT limits provide a tractable setting where the number $N$ of colors can be kept finite, thus allowing to reach a non-perturbative regime in the dual string theory side of the holographic duality.

SMTs have some peculiar features.
Firstly, there exist 12 possible consistent decoupling limits~\cite{Harmark:2007px}.\footnote{See table~\ref{tab:BPS} for the definition of the limits.}
Secondly, each SMT is closed under the action of the one-loop dilatation operator of the parent $\mathcal{N}=4$ SYM theory, and they do not receive contributions from higher loops. 
Therefore, we refer to the possible SMT limits as different \textit{sectors}, since they do not mix with each other.
Finally, SMTs are non-Lorentzian theories.
One manifestation of this feature is that SMTs enjoy an emergent $\mathrm{U}(1)$ symmetry, interpreted as particle number conservation.
Another indication is that SMTs can be obtained by selecting a subset of modes upon performing a Kaluza-Klein decomposition along the three-sphere of $\mathcal{N}=4$ SYM on $\mathbb{R} \times S^3$.
This operation reduces the original QFT to matrix quantum mechanics (QM).

Notably, a counting of the degrees of freedom surviving the decoupling limits reveals that SMTs effectively behave like $(n+1)$--dimensional theories, where $n=0,1,2$ depending on the specific sector.
Therefore, one may expect that it would be possible to re-express the SMTs with $n=1,2$ in terms of local fields, obtaining a standard QFT description with an associated classical action.
When $n=1$, such a description has been derived in Refs.~\cite{Harmark:2019zkn,Baiguera:2020jgy} by locating the theory on $\mathbb{R} \times S^1$. 
When $n=2$, it is harder to find an action formulation and an appropriate geometry acting as a background.
Important steps in this direction have been taken in~\cite{Lambert:2021nol,Smith:2023jjb,Baiguera:2024vlj,Blair:2024aqz,Lei:2025zal}.

\vskip 5mm

In this work, we focus on the SU(1,1) bosonic sector.
This non-Lorentzian theory admits an exotic action formulation.
The quadratic part can be mapped to a $\beta$-$\gamma$ CFT~\cite{Lesage:2002ch}, being linear in the time and space derivatives.
The two-dimensional scalar field is chiral, since the integer modes in the Fourier expansion over the spatial $S^1$ direction are restricted to be non-negative.
This chiral constraint implies that the equal-time commutation relations are not proportional to a Dirac $\delta$ distribution, but rather to a non-trivial function.
For this reason, we say that the theory admits a \textit{semi-local} formulation.
The SU(1,1) SMT sector also shares similarities with warped CFTs, which are non-Lorentzian theories admitting a chiral dilatation symmetry~\cite{Anninos:2008fx,Detournay:2012pc,Hofman:2014loa,Castro:2015csg,Castro:2015uaa,Afshar:2015wjm,Guica:2017lia,Jensen:2017tnb}.

In order to define a QFT description for the SU(1,1) bosonic subsector, we need to integrate in an auxiliary complex scalar field $A$, which is a remnant of the gauge field in the parent $\mathcal{N}=4$ SYM theory.
The propagation of the $A$ field is instantaneous, reminding us of the behavior of gauge fields in Galilean electrodynamics (GED), a non-relativistic version of Maxwell theory \eg see Refs.~\cite{LeBellac:1973unm,Santos,Festuccia:2016caf,Mehra:2021sfx,Bagchi:2022twx}.
The quantum properties of GED have been investigated in~\cite{Chapman:2020vtn}, showing an undesirable feature: an infinite number of marginal deformations need to be included to make the action renormalizable.
Including $\mathcal{N}=2$ supersymmetry in three dimensions improves this situation by restricting the set of marginal deformations, but this class remains infinite-dimensional~\cite{Baiguera:2022cbp}.
One would hope that the non-Abelian and maximally supersymmetric case -- obtained from the null-reduction of $\mathcal{N}=4$ SYM -- could improve the quantum behavior of the theory.
Here, we regard the bosonic SU(1,1) SMT as a toy model to describe a sensible limit of $\mathcal{N}=4$ SYM from a top-down perspective.

\vskip 5mm

Using a semi-local formulation, we show in this paper that the action of this bosonic SMT is classically invariant under the SU(1,1) spacetime symmetry.
We determine the transformations of coordinates and fields which keep the action invariant.
Next, we expand the fields in Fourier modes over the compact $S^1$ direction, recovering the description in terms of an infinite-dimensional QM theory.
Within this framework, we study the quantum corrections of the theory 
using two approaches: a Feynman diagrammatic method that treats fields as commuting variables in the path integral formulation; and time-dependent perturbation theory by working with quantum operators.
These investigations are carried out in sections~\ref{sec:quantum_SU11_QFT} and \ref{sec:quantum_SMT_QM}, as outlined in detail in table~\ref{tab:intro}.
We find that the two methods agree, giving the following results: the theory is finite at all orders in perturbation theory.
More precisely, the quantum corrections to the self-energy vanish at all loops by normal ordering; while the loop corrections to the quartic vertex of the theory are always finite.

\begin{table}[ht]   
\begin{center}   
\begin{tabular}  {|p{54mm}|c|c|c|} \hline  
 & \textbf{Self-energy}  & \textbf{4-point vertex}   \\ \hline \rule{0pt}{4.9ex}
\rule{0pt}{4.9ex} Feynman diagrams  &  Sections~\ref{ssec:1loop_selfenergy} and \ref{ssec:higher_loops_selfenergy}   &  \, Sections~\ref{ssec:1loop_quartic_vertex} and \ref{ssec:higher_loops_selfenergy}   \\
\rule{0pt}{4.9ex} Time-dep. perturbation theory  &  \, Section~\ref{ssec:check_11_scattering} &  Section~\ref{ssec:check_22_scattering}  \\[0.2cm]
\hline
\end{tabular}   
\caption{Computations performed in various sections of the paper.} 
\label{tab:intro}
\end{center}
\end{table}

This analysis may be considered as a forerunner for understanding the field-theoretical formulation of the SMT limits of $\mathcal{N}=4$ SYM, including the quantum aspects.
The most interesting sector corresponds to the SMT preserving the spin group $\mathrm{PSU}(1,2|3)$, since the gravitational dual admits supersymmetric black hole solutions~\cite{Gutowski:2004yv,Kunduri:2006ek}.

\paragraph{Outline.}
The paper is organized as follows.
We begin by reviewing in section~\ref{sec:preliminaries} the construction of Spin Matrix Theories, referring to their origin as decoupling limits of $\mathcal{N}=4$ super Yang-Mills theory.
We define the so-called SU(1,1) bosonic Spin Matrix Theory, presenting its formulation in terms of semi-local fields.
In section~\ref{sec:classical_SU11_SMT}, we discuss the classical symmetries of the action and show that it is off-shell invariant under the SU(1,1) group.
Sections~\ref{sec:quantum_SU11_QFT} and \ref{sec:quantum_SMT_QM} are the main core of this work, since they contain the computation of the perturbative quantum corrections at all orders in perturbation theory.
The former section presents a diagrammatic approach in terms of Feynman diagrams, while the latter uses time-dependent perturbation theory.
We discuss the results and possible future developments in section~\ref{sec:conclusions}.
Appendix~\ref{app:math_tools} contains additional technical details of the computations performed in the main text.

\section{Preliminaries}
\label{sec:preliminaries}

SMTs allow to study finite--$N$ effects in a controlled way within the framework of the AdS/CFT correspondence.
We introduce in section~\ref{ssec:what_is_SMT} SMTs as quantum mechanical models with an appropriate spin and color structure.
Section~\ref{ssec:SMT_limit_N4SYM} relates SMTs to certain decoupling limits of $\mathcal{N}=4$ SYM on $\mathbb{R} \times S^3$ with gauge group $\mathrm{SU}(N)$, where the number $N$ of colors is kept fixed (and possibly finite).
These decoupling limits define subsectors of the theory that do not mix among each other and are composed by a subset of the original fields. 
We focus in section~\ref{ssec:bos_su11} on the so-called bosonic SU(1,1) sector, which is described by a complex scalar field.
With the above procedure, SMTs are defined as quantum mechanical theories composed by infinitely many matrices, \ie the modes on the three-sphere.
However, one can employ the SU(1,1) structure to find a differential representation for the scalar field (see section~\ref{ssec:semilocal_rep_fields}).
We use this (semi--)local description to find an action formulation of the theory in section~\ref{sec:localsu11}.

\subsection{What is Spin Matrix Theory?}
\label{ssec:what_is_SMT}

SMTs are quantum-mechanical models with a Hilbert space constructed out of a set of harmonic oscillators.
The creation operators generating the space of states are built on a representation $R_s$ of a semi-simple Lie group, called the \textit{spin group}, and on the adjoint representation $R_m$ of the group $\mathrm{SU}(N)$, therefore forming an $N \times N$ \textit{matrix} with $N$ the number of colors.
In the following, we will restrict to the case where the excitations of the system are purely bosonic. 
For the fermionic case and other extensions, we refer the reader to the reviews~\cite{Harmark:2014mpa,Oling:2022fft,Baiguera:2023fus}.

We denote the raising operators of a SMT as $(a^{\dagger}_s)^i_{\,\, j}$, where $s \in R_s$, $i= 1, \dots , N$ labels the fundamental representation of $\mathrm{SU}(N)$ and $j=1, \dots , N$ the anti-fundamental representation of $\mathrm{SU}(N)$.
The ladder operators form the algebra of a harmonic oscillator and define a vacuum state $\ket{0}$ such that
\beq
(a_s)^i_{\,\, j} |0 \rangle =0 \, , \qquad
\left[ (a^r)^i_{\,\, j} , (a^{\dagger}_s)^k_{\,\, l}  \right] = 
\delta^r_s \delta^k_j \delta^i_l \, , \qquad
\forall s \in R_s \, , \,\,\,  \forall  i,j \in \lbrace  1, \dots, N \rbrace \, .
\eeq
The Hilbert space $\mathcal{H}'$ is built by acting with the creation operators on the vacuum state. Physical states form the subset $\mathcal{H} \subset \mathcal{H}'$ of singlets under the $R_m$ representation of the color group $\mathrm{SU}(N)$, \ie in other words the physical states $\ket{\phi}$ are annihilated by the following operator:
\beq
(j_0)^{i}_{\,\, j}  \equiv  \sum_{s \in R_s} \left[ (a^{\dagger}_s)^i_{\,\,k} (a^s)^k_{\,\,j} 
- (a^{\dagger}_s)^k_{\,\,j} (a^s)^i_{\,\,k}  \right]  \, , \qquad
(j_0)^{i}_{\,\, j}  \ket{\phi} = 0 \, .
\label{eq:singlet_condition}
\eeq
By imposing the singlet constraint~\eqref{eq:singlet_condition}, we find that the set of states in the Hilbert space $\mathcal{H}$ is spanned by
\beq
\ket{\phi} \sim
\tr \left( a^{\dagger}_{s_1} \dots a^{\dagger}_{s_l}  \right)
\tr \left( a^{\dagger}_{s_{l+1}} \dots \right) \dots
\tr \left( a^{\dagger}_{s_{k+1}} \dots a^{\dagger}_{s_L} \right) |0 \rangle  \, ,
\label{eq:physical_states_SMT}
\eeq
where $L$ is called the length, and the traces are performed over the color indices of the representation $R_m$.

In this context, we consider quartic interactions composed of two creation and two annihilation operators -- thus preserving the total number of particles -- that commute with the generators of the spin and color groups.
The most general interacting Hamiltonian satisfying the above conditions reads
\beq
H_{\rm int}^{\rm bos} =  - \frac{1}{N} \, U^{s' r'}_{s r} : \tr \left( [a^{\dagger}_{s'}, a^s ] [a^{\dagger}_{r'}, a^{r}] \right) :  
\label{eq:dictionary_bos_term_SMT}
\eeq 
where $:$ denotes the normal ordering, and the coefficients $U^{s' r'}_{s r}$ encode the spin structure of the theory.
In subsection~\ref{ssec:bos_su11}, we will determine the form of $U^{s' r'}_{s r}$ for a specific SMT.

So far, we did not make any assumption on the number $N$ of colors, except that it is fixed.
In the \textit{planar limit} $N = \infty$, the multi-trace states in eq.~\eqref{eq:physical_states_SMT} are linearly independent and the Hilbert space is built out of the single-trace states
\beq
| s_1 \dots s_L \rangle = \tr \left( a^{\dagger}_{s_1} \dots a^{\dagger}_{s_L}  \right) | 0 \rangle \, .
\eeq
Due to the cyclicity of the trace, we can interpret the physical system as a periodic spin chain with translational invariance.
Spin chains have been extensively studied from decoupling limits of top-down models within the framework of the AdS/CFT correspondence, \eg see Refs.~\cite{Ammon:2015wua,Minahan:2010js}.
When $N=\infty$, the interacting Hamiltonian acts on the states of the Hilbert space as follows:
\beq
H_{\rm int}  | s_1 s_2 \dots s_L \rangle =
2 \sum_{k=1}^L U^{m n}_{s_k \, s_{k+1}} | s_1 \dots s_{k-1} \, m \, n \, s_{k+2} \dots s_L \rangle \, ,
\label{eq:Ham_int_nparticles_planar}
\eeq
from which we can extract the coefficients $U^{s' r'}_{s r}$.

Taking $N$ large but not infinite enables to study perturbative corrections in $1/N$, corresponding to the splitting or joining of spin chains.
However, one of the striking features of SMTs is that $N$ can be taken to be finite, not necessarily large.
Since the coefficients $U^{s' r'}_{s r}$ fully characterize the interacting Hamiltonian~\eqref{eq:dictionary_bos_term_SMT}, SMTs provide a finite--$N$ extension of spin chains, uniquely determined by the coefficients in eq.~\eqref{eq:Ham_int_nparticles_planar}. 

\subsection{Spin Matrix Theory as a near-BPS limit of $\mathcal{N}=4$ super Yang-Mills}
\label{ssec:SMT_limit_N4SYM}

We previously defined SMT as a generalization of spin chains, starting from a Hilbert space with an appropriate spin and color structure.
In this subsection, we discuss how SMTs can be obtained from a top-down perspective as particular limits of the AdS/CFT correspondence.
Let us consider $\mathcal{N}=4$ SYM with gauge group $\mathrm{SU}(N)$ on $\mathbb{R} \times S^3$.\footnote{In this work, we will set the radius of the three-sphere to unity.}
We denote with $\mathbf{S}_1, \mathbf{S}_2$ are the Cartan generators for rotations, and $\mathbf{Q}_1,\mathbf{Q}_2,\mathbf{Q}_3$ the Cartan generators for the $\mathrm{SU}(4)$ R-symmetry.
We denote with $D$ the dilatation operator, whose loop expansion reads
\beq
D = D_0 + \delta D = D_0 + \lambda D_2 + \dots  \, ,
\label{eq:expansion_dilatation}
\eeq
where $\lambda$ is the 't Hooft coupling, $D_0$ the tree level dilatation operator, $D_2$ the one-loop contribution, and $\delta D$ the full quantum correction.

The dilatation operator is related to the Hamiltonian $H$ (with eigenvalues $E$) on $\mathbb{R} \times S^3$ via the state-operator correspondence.
In this picture, it is natural to decompose the fields in terms of spherical modes on the three-sphere.
This expansion turns the QFT into an equivalent QM governed by infinitely many matrices.

Next, let us consider a unitarity bound 
\beq
E \geq J \, , \qquad
J \equiv  a_1 \mathbf{S}_1+ a_2 \mathbf{S}_2 + b_1 \mathbf{Q}_1+ b_2 \mathbf{Q}_2 + b_3 \mathbf{Q}_3 \, ,
\label{eq:definition_J}
\eeq
where $J$ is a linear combination of Cartan charges with real coefficients
$\lbrace a_1, a_2, b_1, b_2, b_3 \rbrace$.
SMTs are defined as the effective theories arising in the following near-BPS limit 
\beq
E-J \rightarrow 0  \, , \qquad \lambda \rightarrow 0 \, , \qquad
\frac{E-J}{\lambda} \,\, \mathrm{fixed} \, , \qquad
N \,\, \mathrm{fixed} \, ,
\label{eq:micro_canonical_limit}
\eeq
where $\lambda $ is the 't Hooft coupling.
The conditions~\eqref{eq:micro_canonical_limit} define a \textit{decoupling limit}, where only a subset of the modes of $\mathcal{N}=4$ SYM survive.
Pictorially, the near-BPS limit can be understood as a truncation of the full theory where only states very close to a fixed energy $E=J$ are retained, while at the same time sending the coupling constant to zero.

After the limit, the resulting quantum-mechanical theory is described by the effective Hamiltonian
\beq
H_{\rm SMT} = J + \tilde{g}^2 \lim_{\lambda \rightarrow 0} \frac{E-J}{\lambda} = 
J + \tilde{g}^2 H_{\rm int} \, , 
\label{eq:effective_SMT}
\eeq
where $\tilde{g}$ is a coupling constant.
The theory obtained in this way is a SMT, since it is characterized by fields transforming into the adjoint representation of $\mathrm{SU}(N)$ and the representation $R_s$ of a spin group, which is always a subgroup of PSU$(2,2|4)$.
In terms of the dilatation operator~\eqref{eq:expansion_dilatation}, the effective Hamiltonian~\eqref{eq:effective_SMT} of a SMT is given by
\beq
J = D_0 \, , \qquad
H_{\rm int} = D_2 \, .
\label{eq:SMT_dilatation_operator}
\eeq
In other words, SMTs provide a consistent truncation of $\mathcal{N}=4$ SYM that only retains contributions of the dilatation operator up to one loop.

Importantly for the manipulations of this manuscript, we record here that the quantum corrections to the dilatation operator satisfy the commutation relations~\cite{Beisert:2004ry}
\beq
[D_0, D_2] = 0 \, , \qquad
[D_0, \delta D] = 0 \, .
\label{eq:commutators_dilatation}
\eeq
More generally, the anomalous dilatation operator $\delta D$ commutes with all the classical algebra -- including the tree-level dilatation operator, as recorded above.
As a consequence, one also has $[D_2, D_0 - J] =0$, which is used to show that the sectors defined by the near-BPS limits \eqref{eq:micro_canonical_limit} are closed under the action of the one-loop dilatation operator, \ie they do not mix among each other~\cite{Harmark:2007px}.
There exist several consistent decoupling limits arising from the procedure \eqref{eq:micro_canonical_limit}, as listed in table~\ref{tab:BPS}.
The effective Hamiltonians of these models have been extensively studied in Refs.~\cite{Harmark:2019zkn,Baiguera:2020jgy,Baiguera:2020mgk,Baiguera:2021hky,Baiguera:2022pll}.

\begin{table}[ht]   
\begin{center}   
\begin{tabular}  {|c|c|} \hline   
\textbf{Spin group $G_s$}	& \textbf{Combination of Cartan charges $J$}  \\
\hline
$\SU(2)$ &  $\mathbf{Q}_1+\mathbf{Q}_2$ \\ 
$\SU(1|1)$ &  $\frac{2}{3} \mathbf{S}_1+\mathbf{Q}_1 + \frac{2}{3} \le \mathbf{Q}_2 + \mathbf{Q}_3 \ri$ \\ 
$\SU(1|2)$ &  $\frac{1}{2} \mathbf{S}_1+\mathbf{Q}_1 +\mathbf{Q}_2 + \frac{1}{2} \mathbf{Q}_3 $ \\ 
$\SU(2|3)$ &  $\mathbf{Q}_1+\mathbf{Q}_2+\mathbf{Q}_3$ \\ \hline
 $\SU(1,1)$ bosonic  & $\mathbf{S}_1 + \mathbf{Q}_1$   \\
  $\SU(1,1)$ fermionic  & $\mathbf{S}_1 + \frac{2}{3} ( \mathbf{Q}_1 +  \mathbf{Q}_2 + \mathbf{Q}_3 ) $   \\
	   $\SU(1,1|1)$ &  $\mathbf{S}_1 + \mathbf{Q}_1+ \frac{1}{2}(\mathbf{Q}_2+\mathbf{Q}_3)$ \\
	  $\mathrm{PSU}(1,1|2)$  & $\mathbf{S}_1 + \mathbf{Q}_1+\mathbf{Q}_2$   \\ \hline
	  $\SU(1,2)$ &  $\mathbf{S}_1 + \mathbf{S}_2$  \\
	   $\SU(1,2|1)$  & $\mathbf{S}_1 + \mathbf{S}_2 + \frac{1}{2} \mathbf{Q}_1 + \frac{1}{2} \mathbf{Q}_2$   \\
	   $\SU(1,2|2)$  & $\mathbf{S}_1 + \mathbf{S}_2 +  \mathbf{Q}_1$   \\
	  $\mathrm{PSU}(1,2|3)$  &  $\mathbf{S}_1 + \mathbf{S}_2 +  \mathbf{Q}_1 +\mathbf{Q}_2 + \mathbf{Q}_3$   \\ \hline
\end{tabular}
\caption{List of the non-trivial decoupling limits \eqref{eq:micro_canonical_limit}. The horizontal blocks distinguish theories that are effectively 0+1, 1+1, and 2+1 dimensional from a counting of the degrees of freedom. }
\label{tab:BPS}
\end{center}
\end{table}

From a counting of the number of surviving degrees of freedom, one finds that the SMTs can effectively behave as 0+1, 1+1 or 2+1--dimensional theories.
Therefore, it should be possible to re-express the quantum-mechanical SMTs obtained in the near-BPS limit~\eqref{eq:micro_canonical_limit} in terms of 
(possibly local) fields as in a standard QFT.
In this work, we will study the simplest non-trivial model that behaves as a (1+1)--dimensional theory, \ie the SU(1,1) bosonic subsector.

\subsection{Bosonic SU(1,1) Spin Matrix Theory}
\label{ssec:bos_su11}

The field content of the SU(1,1) bosonic sector consists of a complex scalar $(\Phi_n)^i_{\,\, j}$, where $n \in \mathbb{N}$ is a non-negative integer and $i,j$ are color indices.

\paragraph{Classical Hamiltonian.}
One can get a classical Hamiltonian by performing the near-BPS limit~\eqref{eq:micro_canonical_limit} on the classical $\mathcal{N}=4$ SYM theory, getting~\cite{Harmark:2019zkn,Baiguera:2020jgy}
\begin{subequations}
\beq
H = H_0 + \tilde{g}^2 H_{\rm int} \, ,
\eeq
    \begin{equation}
H_0 = S_1+ Q_1=  \sum_{n=0}^\infty \left( n + \frac{1}{2} \right) \tr |\Phi_n|^2\,,
\label{eq:freeham_H0_SU11}
\end{equation}
\begin{equation}
\label{eq:Hint_su11}
  H_\text{int} = \frac{1}{2N} \sum_{l=1}^{\infty}\frac{1}{l}
\tr \le j^{\dagger}_l j_l \ri 
- \frac{1}{2N} \sum_{n= 0}^\infty h(n) \tr([\Phi_{n}^\dagger, \Phi_{n} ]_{\rm M} \, j_0 )
\, ,
\end{equation}
\label{eq:classical_Hsu11}
\end{subequations}
where $h(n) \equiv \sum_{k=1}^n \frac{1}{k}$ are the \textit{harmonic numbers}.
In the above expressions, we introduced the $\SU(N)$ charge density 
\beq
j_l= \sum_{k =0}^{\infty} [\Phi^{\dagger}_{k}, \Phi_{k+l}]_{\rm M} \, , 
\label{eq:charge_density_scalar_su11bos}
\eeq
where $\text{M}$ denotes a matrix commutator.
All the physical configurations have vanishing $\SU(N)$ charge $j_0=0$ due to the Gauss' law on the three-sphere.
The equal-time Dirac brackets between scalar modes read\footnote{We consider Dirac rather than Poisson brackets because the $\mathcal{N}=4$ SYM theory is subject to constraints. We refer to Ref.~\cite{Baiguera:2020jgy} for more details.}
\begin{equation}
\label{eq:Dirac_bracket_scalar2}
\Big\{ (\Phi_n)^i {}_j (t) , (\Phi_{n'}^\dagger)^k {}_l (t) \Big\}_D = i \delta_{n,n'} \delta^i {}_l \delta^k {}_j \, .
\end{equation}
The classical interacting Hamiltonian~\eqref{eq:Hint_su11} is invariant under the $\SU(1,1)$ symmetry, generated by the charges
\begin{equation}
\label{bossu11_su11gens}
L_0 =  \sum_{n=0}^{\infty}  \le n + \frac{1}{2} \ri \tr |\Phi_{n} |^2  \spa
L_+ = (L_-)^{\dagger} = \sum_{n=0}^{\infty} \le n+1 \ri \tr ( \Phi^{\dagger}_{n+1} \Phi_n ) \, ,
\end{equation}
which satisfy the brackets
\beq
\lbrace L_0 , L_{\pm} \rbrace_D = \pm i L_{\pm} \, , \qquad
\lbrace L_+, L_- \rbrace_D = -2i L_0 \, .
\eeq
The Hamiltonian is manifestly invariant under a global $\mathrm{U}(1)$ symmetry, that we interpret as the non-relativistic particle number conservation.

\paragraph{Quantum Hamiltonian.}
At quantum level, classical fields $\Phi_s$ are promoted to operators $a_s$, and the Dirac brackets~\eqref{eq:Dirac_bracket_scalar2} become commutation relations:
\beq
[ (a_r)^i_{\,\, j} , (a^{\dagger}_s)^{k}_{\,\, l}  ] = \delta^i_{\,\, l} \delta^k_{\,\, j} \delta_{rs} \, .
\label{eq:commutation_relations_ladder_operators_SMT}
\eeq
The Gauss' law is replaced by the $\SU(N)$ singlet condition~\eqref{eq:singlet_condition}, which identifies the set of physical states in the Hilbert space.

The quantum SMT Hamiltonian is obtained by directly promoting the classical Hamiltonian~\eqref{eq:classical_Hsu11} at quantum level, without changes of ordering:
\beq
\begin{aligned}
H_{\rm SMT} & =  \sum_{n=0}^{\infty} \le n + \frac{1}{2} \ri 
\tr \left( a^{\dagger}_n a_n \right)  \\
& + \frac{\tilde{g}^2}{2N} \sum_{l=1}^{\infty} \frac{1}{l} \tr \left(  j_l^{\dagger} j_l  \right)  - \frac{\tilde{g}^2}{2N} \sum_{s=0}^{\infty} h(s)  \tr \le : [a^{\dagger}_s, a_s]_{\rm M} :  j_0 \ri  \, ,
 \end{aligned}
 \label{eq:quantized_SMT_Hamiltonian}
\eeq
supplemented by the requirement $j_0 \ket{\phi}=0$ on physical states.
It was shown for all SMTs that the quantization of the classical Hamiltonian obtained from a decoupling limit of $\mathcal{N}=4$ SYM is consistent with the quantization of the dilatation operator in eq.~\eqref{eq:SMT_dilatation_operator}, see Refs.~\cite{Harmark:2019zkn,Baiguera:2020jgy,Baiguera:2022pll}.

Indeed, one can show that the expression~\eqref{eq:quantized_SMT_Hamiltonian} can be equivalently recast in terms of normal-ordered quantities as follows:
\beq
\begin{aligned}
 H_{\rm SMT} & =  \sum_{n=0}^{\infty} \le n + \frac{1}{2} \ri \tr \le : a^{\dagger}_n a_n : \ri  \\
&  + \frac{\tilde{g}^2}{2N} \sum_{l=1}^{\infty} \frac{1}{l} \tr \le : j_l^{\dagger} j_l : \ri   
- \frac{\tilde{g}^2}{2N} \sum_{n= 0}^\infty h(n) \tr(: [a_{n}^\dagger,a_{n}]_{\rm M} \, j_0 :) \, .
\end{aligned} 
\label{eq:quantized_normal_ordered_H_su111}
\eeq
In other words, all the contributions from the normal ordering of operators cancel.
Furthermore, one can manifestly relate the free and interacting parts of the SMT Hamiltonian~\eqref{eq:quantized_normal_ordered_H_su111} to the tree level and one-loop contributions of the dilatation operator of $\mathcal{N}=4$ SYM, restricted to the SU(1,1) bosonic subsector~\cite{Beisert:2004ry,Harmark:2019zkn,Baiguera:2020jgy}.

\subsection{Semi-local representations on fields}
\label{ssec:semilocal_rep_fields}

In the group theory language, the modes of $\mathcal{N}=4$ SYM on $\mathbb{R} \times S^3$ can be identified with certain letters in the singleton representation of the algebra $\mathfrak{psu}(2,2|4)$.
In the SU(1,1) bosonic subsector, the one-particle states surviving the near-BPS limit~\eqref{eq:micro_canonical_limit} are denoted with $\ket{d_1^n Z}$, where $n \in \mathbb{N}$, $Z$ is a scalar and $d_1$ a gauge-invariant component of the covariant derivative~\cite{Harmark:2007px}.
In terms of the creation operators $a_n^{\dagger}$ introduced in subsection~\ref{ssec:what_is_SMT}, we have 
\beq
\ket{d_1^n Z} \equiv a^{\dagger}_n \ket{0} \, .
\label{eq:letters_d1nZ}
\eeq
This is the so-called \textit{oscillator representation}, since the $\SU(1,1)$ group acts on a set of ladder operators.
Acting with the SU(1,1) generators~\eqref{bossu11_su11gens} on the single-particle states defined in eq.~\eqref{eq:letters_d1nZ} and using the following (quantum) commutation relations
\begin{equation}
\label{su11_algebra}
[L_0,L_\pm ] = \pm L_\pm \spa [L_-,L_+]=2L_0 \,,
\end{equation}
one finds
\beq
\label{su111_action}
\begin{aligned}
& L_0 | d_1^n Z \rangle =  \le n+ \frac{1}{2} \ri | d_1^n Z \rangle \, ,
\\
& L_+ | d_1^n Z \rangle =  (n+1) | d_1^{n+1} Z \rangle \, ,
\\
& L_- | d_1^n Z \rangle =  n | d_1^{n-1} Z \rangle  \, .
\end{aligned}
\eeq
This set of identities shows that the states transform in the $j=1/2$ representation of $\SU(1,1)$.

The authors of Ref.~\cite{Baiguera:2020jgy} found a representation of the $\SU(1,1)$ generators in terms of differential operators acting on a local field.
To this aim, we introduce a compact spatial direction $x$ satisfying the periodicity condition $x \sim x + 2 \pi$.
As a consequence, a Fourier expansion of (classical) fields on this (1+1)--dimensional spacetime implies a summation over integer modes.
Therefore, we represent the scalar field $\Phi_n$ as follows:
\begin{equation}
\label{Phi_field}
\Phi (t,x) = \sum_{n=0}^\infty \Phi_n(t) e^{-i \left( n+\frac{1}{2} \right) x} \, , \qquad
\Phi^{\dagger} (t,x) = \sum_{n=0}^\infty \Phi^{\dagger}_n(t) e^{i \left( n+\frac{1}{2} \right) x} \,.
\end{equation}
Importantly, we require that $n$ is a \textit{non-negative} integer, since the letters in the SU(1,1) bosonic sector have $n \in \mathbb{N}$.
Moreover, note that $\Phi^{\dagger}$ is antiperiodic on the circle due to the half-integer momentum in the exponential. 
As we will comment in subsection~\ref{sec:localsu11}, $\Phi^{\dagger}$ shares some features with $\beta$-$\gamma$ ghost fields, and thus has a mixture of bosonic and fermionic characteristics. A consistent differential representation on $\Phi^{\dagger}(t,x)$ of the $\SU(1,1)$ generators is
\begin{equation}
\label{su11_alg_pos}
L_0 = - i \partial_x \, , 
\qquad
L_\pm = e^{\pm i (x-t)} \le - i  \partial_x \pm \frac{1}{2} \ri \,.
\end{equation}
By using the identity~\eqref{eq:Dirac_bracket_scalar2}, we get the following (classical) equal-time Dirac brackets between scalar fields:
\begin{subequations}
\begin{equation}
\lbrace \Phi(t,x),\Phi(t,x') \rbrace_D = 0 \spa
\lbrace \Phi(t,x),\Phi(t,x')^\dagger \rbrace_D = i S_{\frac{1}{2}}(x'-x) \,,
\label{eq:ET_commutators_scalar}
\end{equation}
\begin{equation}
S_j(x) =  \sum_{n=0}^\infty e^{i (n+j) x} \, .
\label{eq:Sj_function}
\end{equation}
\end{subequations}
The distribution $S_{\frac{1}{2}}$ in the right-hand side arises because of the positivity constraint on the integer modes; had we summed also over negative $n$, we would get the usual Dirac $\delta$--distribution that appears in the equal-time brackets of standard QFTs.
The above result suggests that $\Phi(t,x)$ behaves as a \textit{semi-local} field.

\subsection{Semi-local formulation of bosonic SU(1,1) Spin Matrix Theory}
\label{sec:localsu11}

Using the semi-local representation~\eqref{Phi_field} of the scalar field, one can find an action formulation of the SU(1,1) bosonic subsector as a QFT.
To this aim, it is necessary to integrate in an auxiliary complex scalar field $A(t,x)$, whose Fourier series is defined in the following way:
\beq
A(t,x) =  \sum_{n=0}^{\infty} A_n (t) e^{- i n x} \, .
\label{A_field}
\eeq
As discussed in Ref.~\cite{Baiguera:2020jgy}, one can interpret this field as a remnant of the gauge field of the parent $\mathcal{N}=4$ SYM theory.

Given a (1+1)--dimensional spacetime with compact direction $x \sim x + 2 \pi$, we define the classical action of the bosonic SU(1,1) sector as follows:
\begin{subequations}
    \beq
S=  \int dt dx \, \tr \left[ - i \Phi^{\dagger} (\partial_t + \partial_x) \Phi - i A^{\dagger} \partial_x A 
+ \frac{\tilde{g}}{\sqrt{N}} \le A^{\dagger} j + A j^{\dagger} \ri \right] \, ,
\label{eq:local_action_su11bos}
\eeq
\beq
j(t,x) \equiv [\Phi^{\dagger}(t,x) , \Phi(t,x)]_{\rm M} \, ,
\label{eq:semilocal_charge_density}
\eeq
\end{subequations}
where $j(t,x)$ is a semi-local charge density associated with the global $\SU{(N)}$ color symmetry.

We briefly show that the mode expansion of the action~\eqref{eq:local_action_su11bos} matches the Legendre transform of the Hamiltonian~\eqref{eq:classical_Hsu11}.
Plugging the decompositions~\eqref{Phi_field} and \eqref{A_field} inside eq.~\eqref{eq:local_action_su11bos}, we find
\beq
S=  \sum_{n=0}^{\infty} \int dt \, \tr \left[ - i \Phi^{\dagger}_n \partial_t \Phi_n -  \le n+\frac{1}{2} \ri \Phi^{\dagger}_n \Phi_n 
+n \, A^{\dagger}_n A_n + \frac{\tilde{g}}{\sqrt{N}} \,  \le A^{\dagger}_n j_n + A_n j^{\dagger}_n \ri  \right]  \, ,
\eeq
where we rescaled the action by a normalization factor of $\int dx = 2 \pi$, and we identified the modes of the charge density $j_n (t)$ with eq.~\eqref{eq:charge_density_scalar_su11bos}.
Since $A(t,x)$ is non-dynamical, its equations of motion give rise to an algebraic constraint 
\beq
n A_n (t) + \tilde{g} \, j_n (t) = 0 \, .
\eeq
When $n=0,$ this coincides with the $\SU{(N)}$ singlet constraint $j_0 = 0.$
When $n>0,$ the constraint can be solved and inserted into the action to get
\beq
S=  \int dt \, \tr  \left[ - \sum_{n=0}^{\infty}   i \Phi^{\dagger}_n \partial_t \Phi_n - \sum_{n=0}^{\infty}  \le n+\frac{1}{2} \ri \Phi^{\dagger}_n \Phi_n 
 - \frac{\tilde{g}^2}{N} \sum_{l=1}^{\infty} \frac{1}{l} \, j^{\dagger}_l j_l + \frac{\tilde{g}}{\sqrt{N}}  (A_0 + A_0^{\dagger}) j_0 \right]  \, .
\eeq
Upon performing a Legendre transform, one finds eq.~\eqref{eq:classical_Hsu11}.\footnote{The precise identification is completed if we also require that $A_0 = \sum_{k=0}^{\infty} h(k) [\phi^{\dagger}_k, \phi_k]_{\mathrm{M}} $.}

The theory admits various non-Lorentzian features.
One of them is the existence of a global $\mathrm{U}(1)$ symmetry, that we interpret as a non-relativistic particle number conservation.
Another one is that the Fourier expansion on the compact $S^1$ direction is restricted to non-negative modes, giving rise to the non-standard Dirac brackets~\eqref{eq:ET_commutators_scalar}.
The action~\eqref{eq:local_action_su11bos} is also invariant under the global $\mathrm{SU}(N)$ color group, a remnant of the gauge symmetry of the parent $\mathcal{N}=4$ SYM.
We will discuss the invariance of the action under SU(1,1) symmetry in section~\ref{sec:classical_SU11_SMT}.

Finally, we point out a connection between the free part of the SU(1,1) action and the $\beta$-$\gamma$ CFT~\cite{Lesage:2002ch}. 
Upon introducing two real scalar fields $(\beta, \gamma)$ as
\beq
\Phi = \beta + i \gamma  \,,
\label{eq:phi_as_beta_gamma}
\eeq
the scalar kinetic term of the action \eqref{eq:local_action_su11bos} can be recast in the form
\beq
\label{betagammaL}
- i \Phi^{\dagger} (\partial_0 + \partial_x) \Phi = 2 \beta (\partial_0 +\partial_x) \gamma  \, .
\eeq
Notice that it is crucial that the scalar field is complex: both to make sense of the map~\eqref{eq:phi_as_beta_gamma} to the $\beta$-$\gamma$ CFT, and because the kinetic term in the action would be a total derivative if $\Phi$ was real.
Interestingly, the ghost-like behavior of scalar fields in this two-dimensional QFT reminds us of the four-dimensional non-unitary chiral CFT~\cite{Beem:2013sza}.
It would be interesting to relate the SU(1,1) bosonic sector with the chiral algebra. We leave this topic for future studies.

\section{Classical properties of SU(1,1) SMT}
\label{sec:classical_SU11_SMT}

In this section, we discuss the SU(1,1) spacetime symmetry -- corresponding to the spin group defining the near-BPS limit~\eqref{eq:micro_canonical_limit} -- of the semi-local classical action~\eqref{eq:local_action_su11bos}. 
We investigate the off-shell transformations of coordinates and fields in subsection~\ref{ssec:SU11_invariance_offshell}.
We then compute the associated conserved charge in subsection~\ref{ssec:conserved_charge_su11}.
Finally, we analyze the on-shell invariance of the action in subsection~\ref{ssec:onshell_su11}.
In the remainder of this section, we will assume that the Gauss' law is employed, therefore setting $j_0=0$.

\subsection{Off-shell SU(1,1) invariance}
\label{ssec:SU11_invariance_offshell}

The SU(1,1) group is isomorphic to SO(2,1), which can be interpreted as either the one-dimensional global conformal group, or the three-dimensional Lorentz group.
In all the cases, it is evident that we are talking about a spacetime symmetry.
Let us consider the following general transformations on the coordinates and fields
\begin{subequations}
    \beq
\tilde{t} = t - \varepsilon \, a^0 (t,x) \, , \qquad
\tilde{x} = x - \varepsilon \, a^1 (t,x) \, , 
\label{eq:2d_coord_transf}
\eeq
\beq
  \tilde{\Phi}=(1+\varepsilon g(t,x))\Phi \, , \qquad
    \tilde{\Phi}^\dagger=(1+\varepsilon g^*(t,x))\Phi^\dagger \, , 
    \label{eq:transf_scalar_su11}
\eeq
\beq
 \tilde{A}=(1+\varepsilon f(t,x)) A \, , \qquad
    \tilde{A}^\dagger=(1+\varepsilon f^*(t,x)) A^\dagger \, ,
    \label{eq:transf_A_su11}
\eeq
\label{eq:all_su11_transformations}
\end{subequations}
where $\varepsilon$ is an infinitesimal parameter and $a^0, a^1, f, g$ are arbitrary functions that we will determine below.
Our goal is to study the conditions on these functions such that the action~\eqref{eq:local_action_su11bos} is invariant under the transformations~\eqref{eq:all_su11_transformations}, up to $\mathcal{O}(\varepsilon^2)$ terms.
This will provide a proof of the off-shell invariance of the theory under SU(1,1) symmetry.
The non-trivial transformations of the fields are justified by recalling that the scalar field $\Phi$ transforms under the $j=1/2$ representation of the SU(1,1) spin group, as we observed in subsection~\ref{ssec:semilocal_rep_fields}.

First of all, we find that the integration measure transforms as follows under the transformations~\eqref{eq:2d_coord_transf}:
\beq
dt dx = d\tilde{t} d\tilde{x} \left[ 1 + \varepsilon \le \p_0 a^0 + \p_1 a^1 \ri  \right] + \mathcal{O}(\varepsilon^2) \, ,
\label{eq:change_measure}
\eeq
where we used the notation $\p_0 \equiv \p_t$ and $\p_1 \equiv \p_x$.
The infinitesimal variations of the partial derivatives read
\begin{subequations}
\beq
\partial_0  =
\le 1 - \varepsilon \partial_0 a^0 \ri \tilde{\partial}_0  
- \varepsilon (\partial_0 a^1) \tilde{\partial}_1 \, ,
\eeq
\beq
\partial_1  =
- \varepsilon \le  \partial_1 a^0 \ri \tilde{\partial}_0  
+ \le 1 - \varepsilon \partial_1 a^1 \ri \tilde{\partial}_1 \, .
\eeq
\label{eq:transf_partial_der1}
\end{subequations}
For the upcoming manipulations, it is useful to introduce the light-cone coordinates
\beq
x^{\pm} = t \pm x \, , \qquad
\partial_{\pm} = \frac{1}{2} \le \partial_0 \pm \partial_1 \ri \, .
\label{eq:lightcone_coord}
\eeq

\subsubsection{Invariance of the free action}

The structure of eq.~\eqref{eq:local_action_su11bos} implies that the various terms of the action need to be separately invariant under the SU(1,1) transformations.

We begin by considering the kinetic term for the scalar field.
Using the variations~\eqref{eq:all_su11_transformations}, \eqref{eq:change_measure} and \eqref{eq:transf_partial_der1}, we find
\beq
\begin{aligned}
    & \int dt \, dx \, \tr \left\lbrace \Phi^\dagger (\p_0 + \p_1) \Phi \right\rbrace  = \\
    =& \int d \tilde{t} \, d\tilde{x} \, \tr  \left\lbrace \left[ 1 + \varepsilon \p_1 (a^1 - a^0) - \varepsilon (g^*+g)  \right] \tilde{\Phi}^\dagger \tilde{\p}_0 \tilde{\Phi} + \left[ 1 - \varepsilon \p_0 (a^1 - a^0) - \varepsilon (g^*+g) \right] \tilde{\Phi}^\dagger \tilde{\p}_1 \Phi  \right. \\
    & \left. 
    - \varepsilon (\p_0 + \p_1) g \, \tilde{\Phi}^\dagger \tilde{\Phi} \right\rbrace 
    + \mathcal{O}(\varepsilon^2) 
    \, .
\end{aligned}
\eeq
To ensure the invariance of this part of the action, we need to impose the following conditions:
\beq
\begin{cases}
    \p_1 (a^1 - a^0)=g^*+g \, ,
    \\
     \p_0 (a^1 - a^0)=-(g^*+g) \, ,
   \\
     ( \p_0 + \p_1) g=0  \, . 
    \label{eq:constraint3_free_su11}
\end{cases}
\eeq
In terms of the light-cone coordinates~\eqref{eq:lightcone_coord}, the last constraint in eq.~\eqref{eq:constraint3_free_su11} implies that $g$ is only a function of $x^-$.
If we define $a^- \equiv a^0 - a^1$, by summing the first two equations in~\eqref{eq:constraint3_free_su11}, we get
\beq
(\partial_0 + \partial_1) a^- = 0 \, ,
\eeq
meaning that $a^-$ is only a function of $x^-$.

Next, we consider the kinetic term for the auxiliary field $A$.
Upon implementing the coordinate transformations~\eqref{eq:all_su11_transformations}, we obtain
\beq
\begin{aligned}
& \int dt \, dx \, \tr \le A^{\dagger} \partial_1 A \ri  = \\
& = \int d\tilde{t} \, d\tilde{x} \, \tr
\left\lbrace \tilde{A}^{\dagger} \tilde{\partial}_1 \tilde{A} 
- \varepsilon (\partial_1 a^0) \tilde{A}^{\dagger} \tilde{\partial}_0 \tilde{A} - \varepsilon \le f + f^* - \partial_0 a^0 \ri \tilde{A}^{\dagger} \tilde{\partial}_1 \tilde{A} 
- \varepsilon (\partial_1 f) \, \tilde{A}^{\dagger} \tilde{A}
\right\rbrace \, .
\end{aligned}
\eeq
The above expression remains invariant if we impose the following constraints:
\beq
\begin{cases}
      f + f^*- \partial_0 a^0 =0 \, , \\
    \partial_1 a^0 = 0 \, , \\
     \partial_1 f = 0 \, .
\end{cases}
\label{eq:identities_invariance_step2}
\eeq
In other words, $f$ and $a^0$ are only functions of the time coordinate $t$.

The union of the conditions~\eqref{eq:constraint3_free_su11} and \eqref{eq:identities_invariance_step2} keeps the quadratic part of the classical action~\eqref{eq:local_action_su11bos} invariant under infinitesimal SU(1,1) transformations.
In particular, the combination of the above constraints implies
\beq
\p_1 a^1 = g + g^* \, .
\eeq
Therefore, $\partial_1 a^1$ is only a function of $x^-$.

\subsubsection{Invariance of the interacting action}

The interacting part of the classical action~\eqref{eq:local_action_su11bos} mixes the contributions from all the functions $a^0, a^1, f, g$.
First, the variation~\eqref{eq:transf_scalar_su11} of the scalar field implies that the charge density~\eqref{eq:semilocal_charge_density} changes as follows:
\beq
\tilde{j} = (1 + \varepsilon g(t,x)) (1+ \varepsilon g^*(t,x) ) j \, , \qquad
\tilde{j}^{\dagger} = (1 + \varepsilon g(t,x)) (1+ \varepsilon g^*(t,x) ) j^{\dagger}\, . 
\eeq
Therefore, the interacting term reads
\beq
\int dt \, dx \, \tr \left(  A^{\dagger} j 
+ A j^{\dagger} \right)  = \int d\tilde{t} \, d\tilde{x} \, 
\left[  \tilde{A}^{\dagger} \tilde{j} + \varepsilon \tilde{A}^{\dagger}  \left( \partial_0 a^0 + \partial_1 a^1 - f^* - g - g^*  \right) \tilde{j}   \right] \, ,
\eeq
which is invariant if we impose
\beq
\partial_0 a^0 + \partial_1 a^1 = f^* + g + g^* \, . 
\label{eq:invariance_interacting}
\eeq
By plugging the identities~\eqref{eq:constraint3_free_su11} and \eqref{eq:identities_invariance_step2} inside~\eqref{eq:invariance_interacting}, we obtain
\beq
f = f^* = 0 \, .
\eeq
Collecting now all the conditions obtained so far, we obtain the set of constraints that the functions entering eq.~\eqref{eq:all_su11_transformations} need to satisfy:
\beq
\begin{cases}
    \partial_0 a^0 = \partial_1 a^0 = 0 \, , \\
    f = f^* = 0 \, , \\
    \partial_1 a^1 = - \partial_0 a^1 = g + g^* \, , \\ 
    (\p_0 + \p_1) g = 0 \, . \\
\end{cases}
\label{eq:final_constraints_su11_local}
\eeq
In other words, the auxiliary field $A$ is invariant under SU(1,1) transformations, $a^0$ is a constant, and $a^1, g$ are purely functions of $x^-$.

\subsection{Conserved charges of the SU(1,1) symmetry}
\label{ssec:conserved_charge_su11}

Using the constraints~\eqref{eq:final_constraints_su11_local} inside the infinitesimal variations~\eqref{eq:all_su11_transformations}, we finally obtain the general form of the SU(1,1) transformations:
\beq
\begin{aligned}
 & \tilde{t} = t - \varepsilon \, a^0  \, , \qquad
 \tilde{x} = x - \varepsilon \, a^1 (t-x) \, ,  \\
&   \tilde{\Phi}=(1+\varepsilon g(t-x))\Phi \, , \qquad
  \tilde{A}= A  \, .
\label{eq:final_all_su11_transformations}
\end{aligned}
\eeq
Next, we apply Noether's theorem to find the conserved charge associated with this symmetry.
A direct computation gives
\beq
 Q=\int dx \left[ \Phi^\dagger \partial_x \Phi (a^0-a^1)-\Phi^\dagger g \Phi \right] \, . 
\label{eq:charge_semilocal_free}
\eeq
As a consistency check, we show below that the semi-local conserved charge~\eqref{eq:charge_semilocal_free} can be identified with the three conserved charges $L_0, L_{\pm}$ introduced in eq.~\eqref{su11_alg_pos} upon expansion over Fourier modes.

We can assume, without loss of generality, that $g$ is a real function.
The set of equations~\eqref{eq:final_constraints_su11_local} admits the following three linearly independent solutions:
\begin{subequations}
\beq
a^0 - a^1 = \text{const.} \, , \qquad
g = 0 \, , 
\label{eq:solution1_a0a1g}
\eeq
\beq
a^0 - a^1 = \sin (x-t) \, , \qquad
g = \frac{1}{2} \cos (x-t) \, , 
\label{eq:solution2_a0a1g}
\eeq
\beq
a^0 - a^1 = \cos (x-t) \, , \qquad
g = - \frac{1}{2} \sin (x-t) \, .
\label{eq:solution3_a0a1g}
\eeq
\end{subequations}
Let us plug the first solution~\eqref{eq:solution1_a0a1g} inside the charge~\eqref{eq:charge_semilocal_free}, and employ the decomposition~\eqref{Phi_field} of the scalar field in positive modes.
We find
\beq
    Q_1 = -i \sum_{n=0}^{\infty} \le n+\frac{1}{2} \ri {\rm tr} |\Phi_n|^2 \, .
    \label{eq:charge1}
\eeq
Next, we determine the charge associated with the second and third solutions~\eqref{eq:solution2_a0a1g}--\eqref{eq:solution3_a0a1g}, respectively.
After expressing the trigonometric functions in terms of imaginary exponentials, we obtain
\begin{subequations}
\beq
Q_2 = \sum_{n=0}^{\infty} {\rm tr} \left[ -\frac{1}{2}(n+1)\Phi_n^\dagger \Phi_{n+1} e^{-it}+\frac{1}{2}(n+1)\Phi_{n+1}^\dagger \Phi_n e^{it} \right] \, ,
\eeq    
\beq
Q_3 = \sum_{n=0}^{\infty} {\rm tr} \left[ -\frac{i}{2}(n+1)\Phi_n^\dagger \Phi_{n+1} e^{-it}-\frac{i}{2}(n+1)\Phi_{n+1}^\dagger \Phi_n e^{it} \right] \, .
\eeq
\end{subequations}
It is now easy to match the above charges with the SU(1,1) generators in eq.~\eqref{bossu11_su11gens} by taking the following linear combinations:
\begin{gather}
    i Q_1= L_0 \, , \\
    (i Q_3+Q_2) e^{-it}=L_+  \, , \\
    (i Q_3-Q_2) e^{it}=L_- \, .
    \label{eq:dictionary_Q_L_gen}
\end{gather}
This shows that the identities~\eqref{eq:final_all_su11_transformations} correctly parametrize an infinitesimal SU(1,1) transformation for the semi-local representation of the fields in the SMT limit under consideration.

\subsection{On-shell SU(1,1) symmetry}
\label{ssec:onshell_su11}

It is interesting to study the on-shell SU(1,1) invariance of the action after expanding the semi-local fields into Fourier modes over the spatial circle and integrating out the auxiliary field.
For definiteness, we refer to the free and interacting part of the Lagrangian as
\begin{subequations}
\beq
    \mathcal{L} \equiv \mathcal{L}_0 + \tilde{g}^2 \mathcal{L}_{\rm int} \, ,
\eeq
\beq
\mathcal{L}_0 \equiv  \sum_{n=0}^{\infty}  \tr \left[ -i  \Phi^{\dagger}_n \partial_t \Phi_n  -  \le n+\frac{1}{2} \ri \Phi^{\dagger}_n \Phi_n  \right]
\, ,
\eeq
\beq
\mathcal{L}_{\rm int} \equiv  - \frac{1}{2N} \sum_{n=1}^{\infty} \frac{1}{n} \, \tr \left( j^{\dagger}_n j_n \right) \, .
\eeq
\label{eq:Lagrangian_modes_su11}
\end{subequations}
One can compare the Lagrangian~\eqref{eq:Lagrangian_modes_su11} with the Hamiltonian~\eqref{eq:classical_Hsu11}.
Of course, by definition of Legendre transform, we have $\mathcal{L}_{\rm int} = - H_{\rm int}$.

By using the Dirac brackets~\eqref{eq:Dirac_bracket_scalar2}, it can be checked by direct computation that the interacting Lagrangian (and Hamiltonian) are invariant under SU(1,1) transformations.
This statement is equivalent to the vanishing of the following Dirac brackets:
\beq
\lbrace L_0, \mathcal{L}_{\rm int} \rbrace_D = 
\lbrace L_0, H_{\rm int} \rbrace_D = 
\lbrace L_{\pm}, \mathcal{L}_{\rm int} \rbrace_D = 
\lbrace L_{\pm}, H_{\rm int} \rbrace_D = 0 \, .
\eeq
However, the free Hamiltonian is \textit{not} invariant under the full SU(1,1) symmetry, since
\beq
\lbrace L_0, H_0 \rbrace_D =0 \, , \qquad
\lbrace L_{\pm} , H_0 \rbrace_D = \mp i L_{\pm} \, .
\eeq
Notice that this fact is a consequence that the Hamiltonian selects a time direction, thus breaking part of the symmetries.

On the contrary, using the classical equations of motion
\beq
\p_t \Phi_n = \lbrace H_0, \Phi_n \rbrace_D = - i \le n+\frac{1}{2} \ri \Phi_n \, ,
\label{eq:EOM_modes_freeaction}
\eeq
it is easy to check that the free Lagrangian is invariant on-shell:
\beq
\lbrace L_0, \mathcal{L}_0 \rbrace_D  =
\lbrace L_{\pm}, \mathcal{L}_0 \rbrace_D = 0 \, .
\eeq
As a result, we found that the full Lagrangian~\eqref{eq:Lagrangian_modes_su11} has vanishing Dirac brackets with all the generators of the SU(1,1) group.
Of course, this is consistent with the off-shell invariance of the action proven in subsection~\ref{ssec:SU11_invariance_offshell} using the semi-local representation of the SMT.
\textit{En passant}, we notice that the equations of motion~\eqref{eq:EOM_modes_freeaction} are solved by
\beq
\Phi_n (t) = c_n e^{-i \le n+\frac{1}{2} \ri t} \, , \label{eq:on_shell_mode_solution}
\eeq
with coefficients $c_n$ satisfying the Dirac brackets
\beq
\lbrace c_n, c^{\dagger}_m \rbrace_D = i \delta_{mn} \, .
\eeq

\section{Quantum corrections of SU(1,1) SMT (Feynman diagrams)}
\label{sec:quantum_SU11_QFT}

In this section, we study the quantum features of the bosonic SU(1,1) SMT.
Our main goal is to determine the renormalization properties of the theory using Feynman diagrams, in order to make contact with the standard treatment of quantum corrections performed in QFT.
To this aim, it turns out that it is easier to work in momentum space along the compact spatial direction.
The starting point is the classical action in section~\ref{ssec:classical_action}.
We find the corresponding Feynman rules in section~\ref{ssec:Feynman_rules}.
We then employ the Feynman diagrammatic technology to compute the quantum corrections to the effective action. 
Section~\ref{ssec:renormalizability_theory} contains an analysis of the renormalizability of the theory, based on the superficial degree of divergence.
We then compute the one-loop corrections to the self-energy in section~\ref{ssec:1loop_selfenergy} and to the quartic vertex in section~\ref{ssec:1loop_quartic_vertex}.
Higher loops are discussed in section~\ref{ssec:higher_loops_selfenergy}.

\subsection{Classical action}
\label{ssec:classical_action}

A priori, one could evaluate the quantum corrections of the bosonic SU(1,1) SMT in terms of the Feynman diagrams associated with the semi-local formulation presented in eq.~\eqref{eq:local_action_su11bos}.
However, it turns out that this procedure is hard for two main reasons: the theory admits constraints that should be treated with the Dirac method; and the brackets~\eqref{eq:ET_commutators_scalar} are non-standard, thus making harder the determination of propagators and vertices.
We leave this approach for future studies.

Instead, in this manuscript we will explicitly employ the Fourier expansion over the $S^1$ direction of the spacetime, using the action
\beq
\begin{aligned}
S = \int dt \, \tr & \left[ - \sum_{n=0}^{\infty}   i \Phi^{\dagger}_n (t) \partial_t \Phi_n (t) - \sum_{n=0}^{\infty}  \le n+\frac{1}{2} \ri \Phi^{\dagger}_n (t) \Phi_n (t)  \right. \\
& \left.  - \frac{\tilde{g}^2}{2N} \sum_{n=1}^{\infty} \frac{1}{n} j^{\dagger}_n j_n  
 + \frac{\tilde{g}^2}{2 N} \sum_{n=0}^{\infty} h(n) \, [ \Phi^{\dagger}_n, \Phi_n ]_{\rm M} j_0     
 \right]  \, ,
 \end{aligned}
 \label{eq:SU11action_momentum}
\eeq
where we already integrated out the auxiliary field $A$ (see discussion in subsection~\ref{sec:localsu11}), and used the charge density introduced in eq.~\eqref{eq:charge_density_scalar_su11bos}. 
In the above expression, we restored the term proportional to $j_0$ (which vanishes upon applying Gauss' law for the global $\mathrm{SU}(N)$ symmetry).

The theory~\eqref{eq:SU11action_momentum} obtained after the expansion over the spatial direction is a quantum mechanics with an infinite number of positive modes, encoded by the fields $\Phi_n (t)$.
We recall that these objects transform in the adjoint representation of the color group, therefore they can be represented as follows:
    \beq
\Phi^i_{\,\,j} \equiv \Phi^a (T_a)^i_{\,\,j} \, , \qquad
\tr \le T^a_R T^b_R \ri = T(R) \delta^{ab} \, , \qquad
T(\text{fund.}) = \frac{1}{2} \, ,
\label{eq:fields_matrices}
\eeq
where $T_a$ are generators of $\mathrm{SU}(N)$, and $T(R)$ is the index associated with the representation $R$ (see section~70 of Ref.~\cite{Srednicki:2007qs} or section~25 of \cite{Schwartz:2014sze} for more details).\footnote{We stress that while the scalar field transforms in the adjoint representation of $\mathrm{SU}(N)$, implying that $\Phi = \Phi^a T_a$, the generators $T_a$ are $N \times N$ matrices written in their fundamental representation.}
Using these definitions, the action~\eqref{eq:SU11action_momentum} can be put in the form
\beq
\begin{aligned}
S= \frac{1}{2} \int dt \, & \left[ - \sum_{n=0}^{\infty}  (\Phi^{\dagger}_a)_n  \le i \partial_t +   n + \frac{1}{2}  \right) (\Phi_a)_n  \right. \\
& \left.  -  \sum_{n=1}^{\infty} \sum_{k,l=0}^{\infty} \frac{\tilde{g}^2}{2N} \, \frac{1}{n} f^{abe} f^{cde} (\Phi^{\dagger}_a)_{k+n} (\Phi_b)_k (\Phi^{\dagger}_c)_l  (\Phi_d)_{l+n} \right. \\
 & \left.  + \frac{\tilde{g}^2}{2 N} \sum_{k,l=0}^\infty h(k)f^{abe}f^{cde}(\Phi_a^\dagger)_k (\Phi_b)_k (\Phi_c^\dagger)_l  (\Phi_d)_l 
 \right]  \, .
 \end{aligned}
\label{eq:SU11action_after_trace}
\eeq

\subsection{Feynman rules}
\label{ssec:Feynman_rules}

At quantum level, the path integral reads
\beq
\mathcal{Z}[J, J^{\dagger}] = 
\int [\mathcal{D}\Phi \, \mathcal{D} \Phi^{\dagger}] \, 
\exp \left[  i S + i \int dt \, \sum_{n=0}^{\infty} \left( J_n \Phi_n 
+ J^{\dagger}_n \Phi^{\dagger}_n \right) \right] \, ,
\eeq
where $S$ is the classical action in eq.~\eqref{eq:SU11action_after_trace}, and $J_n, J_n^{\dagger}$ are complex scalar modes acting as sources.
Correlation functions are computed from this generating functional by taking functional derivatives defined as follows:
\beq
\frac{\delta J_n(t)}{\delta J_m(t')} = \delta_{mn} \delta (t-t') \, , \qquad
\frac{\delta J^{\dagger}_n(t)}{\delta J^{\dagger}_m(t')} = \delta_{mn} \delta (t-t') \, .
\eeq
We denote the effective action as follows:
\beq
\begin{aligned}
\Gamma [(\Phi_n)_{\rm cl}]  = &
\sum_{k=1}^{\infty} \frac{1}{k!} 
\int dt_1 \dots dt_{2m} \, \Gamma^{(2m)}(t_1, \dots, t_{2m})  \\
& (\Phi_{n_1})_{\rm cl} (t_1) \dots (\Phi_{n_{m}})_{\rm cl} (t_m) 
(\Phi^{\dagger}_{n_{m+1}})_{\rm cl} (t_{m+1}) \dots  (\Phi^{\dagger}_{n_{2m}})_{\rm cl} (t_{2m}) \, ,
\end{aligned}
\label{eq:effective_action}
\eeq
where the subscript \textit{cl} means that the field is classical.
Notice that due to $\mathrm{U}(1)$ particle number conservation, any contribution to the effective actions contains the same number of $\Phi$ and $\Phi^{\dagger}$ fields.

Since time direction is continuous and runs through the range $t \in [-\infty, \infty]$ as in any standard QFT, we can read the Feynman rules directly from the classical action, as usual.
However, we stress that any field of the theory (either physical or virtual) is subject to the constraint that its mode decomposition is restricted to non-negative integers.

By inverting the kinetic term, we obtain the propagator:
\beq
(D_{ab})_{n n'} (t)  = \delta_{ab} \delta_{nn'}  \int_{-\infty}^{\infty} \frac{d\omega}{2\pi}  \frac{- i e^{-i \omega t }}{\omega +  n + \frac{1}{2}  + i \varepsilon} = 
-  \delta_{ab} \delta_{nn'}  \Theta(t) e^{i t \le n + \frac{1}{2} \ri} \, ,
\label{eq:propagator_modes}
\eeq
where $\Theta$ is the Heaviside step distribution.
The non-Lorentzian features of the theory give rise to a peculiar structure of the propagator.
Upon quantization, since particles and antiparticles are separately conserved, the fields $\Phi_n$ act purely as annihilation operators, while $\Phi^{\dagger}_n$ as creation ones.
As a consequence, the propagator is purely retarded and non-vanishing only for positive time separation.
These aspects are similar to the Schr\"{o}dinger propagator in non-relativistic theories, \eg see Refs.~\cite{Bergman:1991hf,Klose:2006dd,Baiguera:2023fus}.

Let us now write the Feynman rules in Fourier space (both along the temporal and spatial directions):
\begin{itemize}
    \item Propagator 
    \beq
\includegraphics[valign=c]{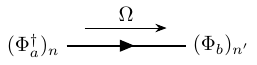}
=
\langle (\Phi_b)_{n'} (\Omega) (\Phi^{\dagger}_a)_{n} (-\Omega) \rangle = \frac{-i \delta_{ab} \delta_{nn'}}{\Omega +  n +\frac{1}{2}  + i \varepsilon} \, .
\label{eq:scalar_propagator_modes}
\eeq
The arrow on the propagator line denotes the $\mathrm{U}(1)$ charge.
\item Quartic vertex 1
\beq
 \mathcal{V}^{n}_{4} \equiv \includegraphics[valign=c]{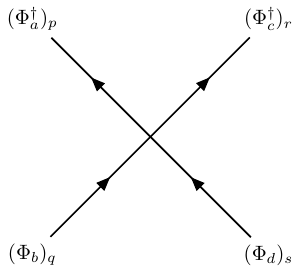}
=
\begin{aligned}
- \frac{i \tilde{g}^2}{2N} \, \frac{1}{n} & \left[ f^{abe} f^{cde} (\delta_{p, q+n}  \delta_{s,r+n}+\delta_{q, p+n}  \delta_{r,s+n})  \right. \\
& \left. -f^{ade} f^{bce} (\delta_{p, s+n}  \delta_{q,r+n}+\delta_{s, p+n}  \delta_{r,q+n}) \right]
\end{aligned}
\label{eq:FR_vertexV4}
\eeq
This vertex carries an additional superscript $n$ over which the interacting term in the action is summed over.
Summing all the above vertices over $n$, we get the following tree-level contribution:
\begin{equation}
  \mathcal{V}_4 \equiv  \sum_{n=1}^{\infty} \mathcal{V}_4^n = 
    - \frac{i \tilde{g}^2}{2N} \delta_{p+r,s+q} 
    \left( f^{abe} f^{cde} \, \frac{1}{|p-q|}  
    - f^{ade}  f^{bce} \, \frac{1}{|p-s|} \right) \, .
\end{equation}
\item Quartic vertex 2
\beq
\tilde{\mathcal{V}}_4 \equiv  \includegraphics[valign=c]{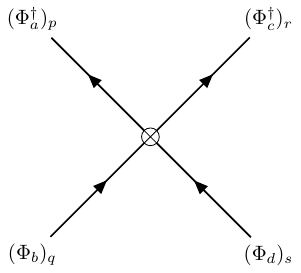}
=
\begin{aligned}
\frac{i \tilde{g}^2}{2N} & \left[ f^{abe} f^{cde} (\delta_{p, q}  \delta_{s,r} h(p) +\delta_{p, q}  \delta_{s,r} h(r) )  \right. \\
& \left. -f^{ade} f^{bce} (\delta_{p, s}  \delta_{q,r} h(p) +\delta_{p, s}  \delta_{q,r} h(r))
\right] 
\end{aligned}
\label{eq:FR_vertexV4tilde}
\eeq
In contrast to $\mathcal{V}_4^n$, this second vertex does \textit{not} carry an additional label $n$, since there are no additional summations in the action~\eqref{eq:SU11action_after_trace}.
\end{itemize}

Notice that the second vertex~\eqref{eq:FR_vertexV4tilde} comes from a contribution in the action proportional to $j_0$, which vanishes on physical states in the Hilbert space.
However, since loop corrections involve virtual processes, we keep it during the computation of Feynman diagrams.
Finally, we point out that the $\mathrm{U}(1)$ charge is conserved at each vertex, as denoted by the same number of ingoing and outgoing arrows.

\subsection{Renormalizability of the theory}
\label{ssec:renormalizability_theory}

In this subsection, we analyze the renormalizability of the theory~\eqref{eq:SU11action_after_trace} by computing the superficial degree of divergence of a general Feynman diagram.
Let us denote with $P$ the number of propagators, $V$ the vertices, $L$ the loops, and $E=E_{S} + E_{\bar{S}}$ the external lines.
The subscripts $S$ ($\bar{S}$) refer to the scalar fields $\Phi$ and their hermitian conjugates $\Phi^{\dagger}$, respectively.

Despite containing an infinite number of fields labeled by a positive integer, the renormalization of the theory~\eqref{eq:SU11action_after_trace} works in the same way as standard QFTs: one needs to introduce local counterterms that get rid of divergences in the UV cutoff, and coupling constants can run if the renormalization procedure forces us to introduce a length scale in the theory.\footnote{A similar analysis can be performed to study the renormalization of 
$\mathcal{N}=4$ SYM on $\mathbb{R} \times S^3$ after expanding the fields into spherical harmonics, \eg see section~6 of Ref.~\cite{Ishiki:2006rt}.}

In the following, we list the constraints on the above data coming from the structure of Feynman rules derived in subsection~\ref{ssec:Feynman_rules}:
\begin{itemize}
    \item Given a connected graph, the topological constraint fixes
\beq
L = P - V + 1 \, .
\label{eq:topological_constraint}
\eeq
\item Each vertex connects exactly 2 scalars $S$ and two hermitian conjugates $\bar{S}$. Similarly, each propagator joins one scalar $S$ with one $\bar{S}$.
As a consequence of these observations, we obtain:
\beq
2V = P + E_{S} \, , \qquad
2V = P + E_{\bar{S}} \, .
\label{eq:scalarsS_conjugates_Sbar}
\eeq
\end{itemize}
Combining the above constraints, we obtain the following set of identities:
\begin{subequations}
\beq
E_{S} = E_{\bar{S}} \, , \qquad
E \in 2 \mathbb{N} \, .
\label{eq:constraints1_Feynman}
\eeq
    \beq
P = \frac{E}{2} + 2L -2 \, , \qquad
V = \frac{E}{2} +L -1   \, .
\label{eq:constraints2_Feynman}
\eeq
\end{subequations}
The conditions~\eqref{eq:constraints1_Feynman} arise from the particle number conservation of non-relativistic theories.
In particular, there are no perturbative quantum corrections to any Feynman diagram with an odd number of external lines.
The identities~\eqref{eq:constraints2_Feynman} allow to express the superficial degree of divergence in terms of the variables $(E,L)$ only.

Next, notice that the Lagrangian~\eqref{eq:SU11action_after_trace} is a continuous function of the $t$ coordinate, while the spatial coordinate was expressed via a discrete summation over positive Fourier modes.
For this reason, we analyze the superficial degree of divergence
associated with the temporal and spatial momenta separately.

\paragraph{Temporal degree of divergence.}
Without loss of generality, we start by analyzing the properties of the integrals over the temporal loop momenta, denoted with $\omega_1, \dots, \omega_L$.
Schematically, the general structure is given by
\beq
\int_{-\infty}^{\infty} d \omega_1 \dots 
\int_{-\infty}^{\infty} d \omega_L \, \frac{1}{(K (\omega_i, p_i))^P} = 
\int_{-\infty}^{\infty} d \omega_1 \dots 
\int_{-\infty}^{\infty} d \omega_L \, \frac{1}{(K (\omega_i, p_i))^{\frac{E}{2}+2L -2}} 
\, ,
\label{eq:temporal_integral_general}
\eeq
where $K(\omega_i, p_i) \equiv \omega_i + p_i + \frac{1}{2} + i \varepsilon$ is the kinetic term in momentum space, and in the second step we used eq.~\eqref{eq:constraints2_Feynman}.
The superficial (temporal) degree of divergence of this expression is defined by
\beq
\Delta_{\omega} \equiv 2 -L - \frac{E}{2} \, ,
\label{eq:temporal_superficial_DOD}
\eeq
\ie it is the difference between the factors of $\omega$ at the numerator (including the measure) minus the ones at the denominator.
The constraints~\eqref{eq:constraints1_Feynman} imply that $E$ is even, meaning that $E \geq 2$.
Moreover, any perturbative quantum corrections includes $L \geq 1$ loops.
Plugging these inequalities inside eq.~\eqref{eq:temporal_superficial_DOD}, we find that 
\beq
\begin{aligned}
& \forall L \geq 1, E \qquad
\Delta_{\omega} \leq 0 \, ,  \\
& L=1, \, E=2 \quad \Leftrightarrow \quad  \Delta_{\omega}=0 \, .
\end{aligned}
\eeq
Therefore, the integral~\eqref{eq:temporal_integral_general} is always convergent except for the one-loop correction to the self-energy, which might be divergent.
Nonetheless, we will discuss around eq.~\eqref{eq:result_integral_1loop_selfenergy_modes} that the corresponding integral can be regularized to 0.
In conclusion, all the energy integrals are convergent, and no renormalization is needed.

\paragraph{Spatial degree of divergence.}
We begin the study of the superficial degree of divergence along the spatial directions by assuming that all the vertices in a Feynman diagram are of the kind~\eqref{eq:FR_vertexV4} -- the other cases will be considered below.
Schematically, the contributions to the Feynman diagrams read
\beq
\begin{aligned}
& \le \sum_{m_1=1}^{\infty} \dots \sum_{m_V=1}^{\infty} \ri
\le \sum_{k_1, l_1=0}^{\infty} \dots \sum_{k_V, l_V=0}^{\infty} \ri
\le \sum_{k'_1, l'_1=0}^{\infty} \dots \sum_{k'_V, l'_V=0}^{\infty} \ri_{\rm constr.}
\frac{1}{(m_i)^V} \,  \delta^{P+2V} = \\
& = \le \sum_{m_1=1}^{\infty} \dots \sum_{m_V=1}^{\infty} \ri
\le \sum_{k_1, l_1=0}^{\infty} \dots \sum_{k_V, l_V=0}^{\infty} \ri
\le \sum_{k'_1, l'_1=0}^{\infty} \dots \sum_{k'_V, l'_V=0}^{\infty} \ri_{\rm constr.}
\frac{1}{(m_i)^V} \,  \delta^{4V-\frac{E}{2}} \, .
\end{aligned}
\label{eq:spatial_summation_general}
\eeq
where we used eq.~\eqref{eq:constraints2_Feynman}, and we denoted with $\delta$ the Kronecker delta, coming from propagators (one from each) and quartic vertices (two from each).

Each vertex of kind~\eqref{eq:FR_vertexV4} brings five summations: over the two incoming momenta $(l_i, k_i)$, the two outcoming momenta $(k'_i, l'_i)$, and the relative difference labelled by $m_i$.
The label \textit{constr.} reminds us that we do not need to sum over the external momenta.

Without loss of generality, we use the Kronecker $\delta$ to get rid of all the possible summations, keeping only one factor of overall momentum conservation.
After this step, the total number of remaining summations is
\beq
\text{$\#$ summations} = \le 5V -E \ri - \le 4V - \frac{E}{2} -1 \ri = 
V - \frac{E}{2} +1 = L \, .
\eeq
At this point, we can evaluate the superficial degree of divergence.
Roughly speaking, in the case of a discrete summation the power counting is analogous to the continuum case upon using the following dictionary:
\beq
\sum_{n \in \mathbb{N}} \sim \sum_{n \in \mathbb{Z}} \sim
\int dn \, .
\eeq
In other words, each summation contribute to a power of 1 in the discrete momenta at the numerator.
By computing the difference between the number of summations and the factors in the spatial momenta $m_i$ (at the numerator with positive sign, at the denominator with negative one) in eq.~\eqref{eq:spatial_summation_general}, we find the spatial degree of divergence $\Delta_p$ as follows:
\beq
\Delta_{p} \equiv (\text{$\#$ summations}) - V =
1 - \frac{E}{2} \, ,
\label{eq:spatial_DOD}
\eeq
From the expression~\eqref{eq:spatial_DOD}, we find that
\beq
\begin{aligned}
& \forall  E \qquad
\Delta_{p} \leq 0 \, ,  \\
&  E=2 \quad \Leftrightarrow \quad  \Delta_{p}=0 \, .
\end{aligned}
\eeq
Therefore, the summations~\eqref{eq:spatial_summation_general} are always convergent except for the computation of the self-energy.
Nonetheless, we will later prove the selection rule~\ref{sel_rule1} according to which all the quantum corrections to the self-energy vanish.
The convergence improves whenever a vertex of kind~\eqref{eq:FR_vertexV4} is replaced by one of kind~\eqref{eq:FR_vertexV4tilde}.

In summary, we have shown by a counting argument that the temporal and spatial degrees of divergence of any connected Feynman diagram associated with the rules in subsection~\ref{ssec:Feynman_rules} are convergent.
The only non-trivial case corresponds to the self-energy, that we will argue to be vanishing in subsections~\ref{ssec:1loop_selfenergy} and \ref{ssec:higher_loops_selfenergy}.
As a consequence, no renormalization will be needed for the theory  with action~\eqref{eq:SU11action_after_trace}.

\subsection{One-loop corrections to the self-energy}
\label{ssec:1loop_selfenergy}

\begin{figure}[ht]
    \centering
   \subfigure[]{\label{subfig:diagram1_1loop_selfenergy_scalarmodes} \includegraphics[width=0.45\linewidth]{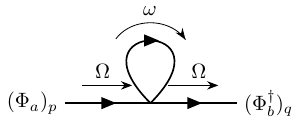}}
   \qquad
     \subfigure[]{\label{subfig:diagram1_1loop_selfenergy_scalarmodes_new} \includegraphics[width=0.45\linewidth]{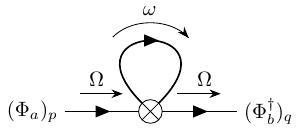}}
    \caption{One-loop contributions to the self-energy of the scalar modes.}
    \label{fig:1loop_selfenergy_tot}
\end{figure}

We consider the one-loop corrections to the self-energy of the scalar modes, defined as $\Gamma^{(2)}$ in eq.~\eqref{eq:effective_action}.
In Fourier space, we decompose the self-energy for convenience as follows:
\beq
i \Gamma^{(2)} = 
(\Phi_a)_p (\Omega) \, (\Phi^{\dagger}_b)_q (\Omega) \,\, \mathcal{I}^{(2)}\, .
\label{eq:decomposition_selfenergy_I}
\eeq
In this subsection, our goal is to compute $\mathcal{I}^{(2)}$.
The Feynman diagrams corresponding to the one-loop corrections are depicted in fig.~\ref{fig:1loop_selfenergy_tot}.

The contribution from the diagram in fig.~\ref{subfig:diagram1_1loop_selfenergy_scalarmodes} reads
\beq
\begin{aligned}
\mathcal{I}^{(2)}_{\ref{subfig:diagram1_1loop_selfenergy_scalarmodes}} &  = -  
\sum_{n=1}^{\infty} \sum_{k,l=0}^{\infty}  \,
\int_{- \infty}^{\infty} \frac{d\omega}{2 \pi} \, \frac{\delta_{k,l} \delta_{cd}}{\omega + k + \frac{1}{2} + i \varepsilon}  \\
& \times 
\frac{\tilde{g}^2}{2 N n} \left[ f^{cae}  f^{bde}  \le \delta_{k,p+n} \delta_{l,q+n} + \delta_{p,k+n} \delta_{q,l+n} \ri
- f^{cde}  f^{abe}  \left( \delta_{k,l+n} \delta_{p,q+n} + \delta_{l,k+n} \delta_{q,p+n} \right)
 \right]   \, .
\end{aligned}
\label{eq:intermediate_step_1loop_selfenergy_scalarmodes3}
\eeq
By solving the constraints given by the Kronecker $\delta$ in each term and using the identities~\eqref{eq:fields_matrices}, the result simplifies to
\beq
\begin{aligned}
 \mathcal{I}^{(2)}_{\ref{subfig:diagram1_1loop_selfenergy_scalarmodes}}   = 
\frac{\tilde{g}^2}{2n}  \delta^{ab}
& \le \sum_{n=1}^{\infty} \sum_{k= \max (n,p)}^{\infty}  \int_{- \infty}^{\infty} \frac{d\omega}{2 \pi} \, \frac{ \delta_{k,p+n} \delta_{k,q+n} }{\omega + n + p + \frac{1}{2} + i \varepsilon} \right. \\
& \left. +  \sum_{n=1}^{p} \sum_{k= 0}^{p}
\int_{- \infty}^{\infty} \frac{d\omega}{2 \pi} \, \frac{\delta_{p,k+n} \delta_{q,k+n}}{\omega + n - p + \frac{1}{2} + i \varepsilon}  \ri  \, ,
\end{aligned}
\label{eq:intermediate_step_1loop_selfenergy_scalarmodes3.2}
\eeq
where we used the antisymmetry of the structure constants to get rid of the terms proportional to $\delta_{cd} f^{cde}$.
The positivity of the integer modes is responsible for the peculiar ranges of the above summations.

Let us evaluate the integrals over the loop momentum $\omega$.
To this aim, we notice that the integral has the same form as the expression entering the tree-level propagator~\eqref{eq:propagator_modes}, in the limit $t \rightarrow 0$.
Therefore, we obtain
\beq
\int_{- \infty}^{\infty} \frac{d\omega}{2 \pi} \, \frac{1}{\omega  + n \pm p + \frac{1}{2} + i \varepsilon}  = 
\lim_{t \rightarrow 0} 
\int_{- \infty}^{\infty} \frac{d\omega}{2 \pi} \, \frac{e^{-i \omega t}}{\omega + n \pm p + \frac{1}{2} + i \varepsilon} =
 -i \Theta(0)  \, .
\label{eq:result_integral_1loop_selfenergy_modes}
\eeq
This result requires a prescription, because the value of the Heaviside distribution is ambiguous at the origin.
In the following, we set $\Theta(0)=0$ because the integral needs to be dimensionless, but it depends on dimensionful quantities.
The same prescription is also adopted in Refs.~\cite{Bergman:1991hf,Klose:2006dd}.
With this choice, we find $\mathcal{I}_{\ref{subfig:diagram1_1loop_selfenergy_scalarmodes}}^{(2)}=0$.

Next, we have a contribution coming from the vertex depicted in Fig.~\ref{subfig:diagram1_1loop_selfenergy_scalarmodes_new}, which reads
\beq
\begin{aligned}
\mathcal{I}^{(2)}_{\ref{subfig:diagram1_1loop_selfenergy_scalarmodes_new}} &  =  \sum_{k,l=0}^{\infty}  \, \int_{- \infty}^{\infty} \frac{d\omega}{2 \pi} \, \frac{\delta_{k,l} \delta_{cd}}{\omega + k + \frac{1}{2} + i \varepsilon} \\
& \times  
\frac{\tilde{g}^2}{2 N} \left[ f^{cae}  f^{bde}  \le h(k) \delta_{k,p} \delta_{l,q} + h(q) \delta_{p,k} \delta_{q,l} \ri
- f^{cde}  f^{abe}  \left( h(k) \delta_{k,l} \delta_{p,q} + h(q) \delta_{k,l} \delta_{p,q} \right)
 \right]  \, .
\end{aligned}
\label{eq:intermediate_step_1loop_selfenergy_scalarmodes_new1}
\eeq
By using the result~\eqref{eq:result_integral_1loop_selfenergy_modes} with the same prescription $\Theta(0)=0$, we also find $\mathcal{I}^{(2)}_{\ref{subfig:diagram1_1loop_selfenergy_scalarmodes_new}}  =0$.
In conclusion, we find that the one-loop correction to the self-energy vanishes:
\beq
\mathcal{I}^{(2)}_{\rm 1-loop} = 0 \, .
\label{eq:total_sum_1loop_selfenergy}
\eeq

\subsection{One-loop corrections to the quartic vertex}
\label{ssec:1loop_quartic_vertex}

\begin{figure}[ht]
    \centering
   \subfigure[] { \label{subfig:1loop_4pt_old_old} \includegraphics[width=0.45\linewidth]{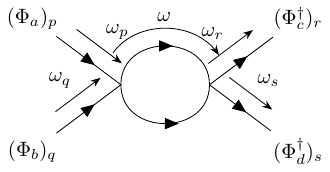}} \quad
   \subfigure[] { \label{subfig:1loop_4pt_new_old} \includegraphics[width=0.45\linewidth]{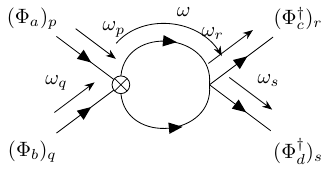}} \\
     \subfigure[] { \label{subfig:1loop_4pt_old_new} \includegraphics[width=0.45\linewidth]{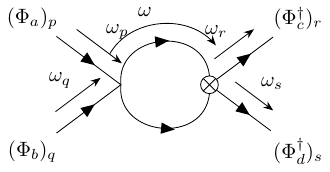}} \quad
   \subfigure[] { \label{subfig:1loop_4pt_new_new} \includegraphics[width=0.45\linewidth]{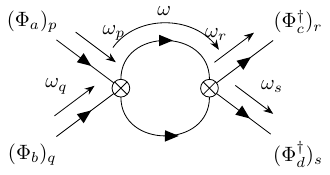}} 
    \caption{One-loop contributions to the four-point vertex of the scalar modes.}
    \label{fig:1loop_4pt}
\end{figure}

We consider the one-loop corrections to the four-point vertex, whose corresponding Feynman graphs are depicted in Fig.~\ref{fig:1loop_4pt}. 
For convenience, we express the contributions to the effective action in the following way:
\beq
i \Gamma^{(4)} = (\Phi_a)_p (\omega_p) (\Phi_b)_q (\omega_q) 
(\Phi^{\dagger}_c)_r (\omega_r) (\Phi^{\dagger}_d)_s (\omega_s)  \, \mathcal{I}^{(4)} \, .
\label{eq:def_quartic_vertex_from_effective_action}
\eeq

It is a tedious exercise to evaluate explicitly all the Feynman diagrams.
We collect the technical details in appendix~\ref{app:ssec:details_1loop_4pt_vertex}, and report here the main steps.
Using the Feynman rules derived in subsection~\ref{ssec:Feynman_rules}, we obtain the following four contributions:
\begin{itemize}[leftmargin=*]
    \item Diagram in fig.~\ref{subfig:1loop_4pt_old_old}:
\beq
\begin{aligned}
 & \mathcal{I}^{(4)}_{\ref{subfig:1loop_4pt_old_old}}  = 
 \sum_{m,n=1}^{\infty} \sum_{i,j,k,l=0}^{\infty}  \le - \frac{i \tilde{g}^2}{2 N m}  \ri  \le  - \frac{i \tilde{g}^2}{2 N n}  \ri   \\
&  \times
\left[   f^{fbv} f^{eav}   \le  \delta_{j, q+m} \delta_{p, i+m} + \delta_{q, j+m} \delta_{i,p+m}  \ri
- f^{fav}  f^{bev}  \le  \delta_{j, p+m} \delta_{q,i+m} 
+ \delta_{p,j+m} \delta_{i,q+m}  \ri \right]  \\
& \times     
\left[  f^{cgw}  f^{dhw}   \le  \delta_{r,k+n} \delta_{l,s+n} + \delta_{k,r+n} \delta_{s,l+n}  \ri
- f^{chw}  f^{gdw}  \le  \delta_{r, l+n} \delta_{k,s+n} 
+ \delta_{l,r+n} \delta_{s,k+n}  \ri \right] \\
& \times
\int_{- \infty}^{\infty} \frac{d\omega}{2 \pi} \, \frac{\delta_{i,k} \delta_{eg}}{\omega + k + \frac{1}{2} + i \varepsilon} 
\frac{\delta_{j,l} \delta_{fh} }{\omega_p+\omega_q-\omega + l + \frac{1}{2} + i \varepsilon} \, . 
\label{eq:1loop_4pt_contr1}
\end{aligned}
\eeq
\item Diagram in fig.~\ref{subfig:1loop_4pt_new_old}:
\beq
\begin{aligned}
& \mathcal{I}^{(4)}_{\ref{subfig:1loop_4pt_new_old}} =  
\sum_{m=1}^{\infty} \sum_{i,j,k,l=0}^{\infty} 
\le \frac{i \tilde{g}^2}{2N} \ri  
\le -\frac{i\tilde{g}^2}{2 N m} \ri 
\\
& \times \left[   f^{fbv} f^{eav}  \delta_{q,j}  \delta_{i,p} \le h(q) + h(p) \ri  
- f^{fav}  f^{bev} \delta_{q,i}  \delta_{j,p} \le h(q) + h(p) \ri  \right]  \\
& \times  \left[ f^{cgw}  f^{dhw} \le  \delta_{k, r+m} \delta_{s, l+m} + \delta_{r, k+m} \delta_{l,s+m}  \ri   
-f^{chw}  f^{gdw} \le  \delta_{k, s+m} \delta_{r, l+m} + \delta_{s, k+m} \delta_{l,r+m}  \ri \right] \\
& \times \int_{- \infty}^{\infty} \frac{d\omega}{2 \pi} \, \frac{\delta_{i,k} \delta_{eg}}{\omega + k + \frac{1}{2} + i \varepsilon} 
\frac{\delta_{j,l} \delta_{fh} }{\omega_p+\omega_q-\omega + l + \frac{1}{2} + i \varepsilon} \, .
\label{eq:1loop_4pt_contr2}
\end{aligned}
\eeq
\item Diagram in fig.~\ref{subfig:1loop_4pt_old_new}:
\beq
\begin{aligned}
& \mathcal{I}^{(4)}_{\ref{subfig:1loop_4pt_old_new}} =  
\sum_{m=1}^{\infty} \sum_{i,j,k,l=0}^{\infty}
\le - \frac{i \tilde{g}^2}{2 N m}  \ri  
\le \frac{i \tilde{g}^2}{2 N} \ri
\\
&  \times   
\left[   f^{fbv} f^{eav}   \le  \delta_{j, q+m} \delta_{p, i+m} + \delta_{q, j+m} \delta_{i,p+m}  \ri
- f^{fav}  f^{bev}  \le  \delta_{j, p+m} \delta_{q,i+m} 
+ \delta_{p,j+m} \delta_{i,q+m}  \ri \right]  \\
& \times   \left[   f^{cgw} f^{dhw}  \delta_{r,k}  \delta_{l,s} \le h(r) + h(s) \ri  
- f^{chw}  f^{gdw} \delta_{r,l}  \delta_{k,s} \le h(r) + h(s) \ri  \right]  \\
& \times \int_{- \infty}^{\infty} \frac{d\omega}{2 \pi} \, \frac{\delta_{i,k} \delta_{eg}}{\omega + k + \frac{1}{2} + i \varepsilon} 
\frac{\delta_{j,l} \delta_{fh} }{\omega_p+\omega_q-\omega + l + \frac{1}{2} + i \varepsilon} \, .
\label{eq:1loop_4pt_contr3}
\end{aligned}
\eeq
\item Diagram in fig.~\ref{subfig:1loop_4pt_new_new}:
\beq
\begin{aligned}
    & \mathcal{I}^{(4)}_{\ref{subfig:1loop_4pt_new_new}}= 
\sum_{i,j,k,l=0}^{\infty} 
\le \frac{i\tilde{g}^2}{2 N} \ri  
\le \frac{i\tilde{g}^2}{2 N} \ri 
\\
& \times
\left[ f^{fbv} f^{eav}  \delta_{j,q}  \delta_{p,i} \le h(p) + h(q) \ri 
- f^{fav} f^{bev} \delta_{p,j} \delta_{q,i} \le h(p) + h(q) \ri \right] \\
& \times  \left[ f^{cgw}  f^{dhw} \delta_{k,r}  \delta_{l,s}  \le h(r) + h(s) \ri
-f^{chw}  f^{gdw}  \delta_{r,l}  \delta_{k,s} \le h(r) + h(s) \ri \right] \\
& \times  \int_{- \infty}^{\infty} \frac{d\omega}{2 \pi} \, \frac{\delta_{i,k} \delta_{eg}}{\omega + k + \frac{1}{2} + i \varepsilon} 
\frac{\delta_{j,l} \delta_{fh} }{\omega_p+\omega_q-\omega + l + \frac{1}{2} + i \varepsilon} \, .
\end{aligned}
\label{eq:1loop_4pt_contr4}
\eeq
\end{itemize}
All the above expressions involve the same convergent integral over the loop momentum $\omega$, which can be solved by means of the residue theorem:
\beq
\int_{- \infty}^{\infty} \frac{d\omega}{2 \pi} \, \frac{1}{\omega + k + \frac{1}{2} + i \varepsilon} 
\frac{1}{\omega_p+\omega_q-\omega + l + \frac{1}{2} + i \varepsilon}  = - \frac{i}{\omega_p + \omega_q + k + l +1 } \, .
\label{eq:integral_4pt_residue}
\eeq
After using the result~\eqref{eq:integral_4pt_residue}, performing the summations over integer momenta as outlined in appendix~\ref{app:ssec:details_1loop_4pt_vertex} and summing the contributions from all the four diagrams, we obtain
\beq
\mathcal{I}^{(4)}_{\rm 1-loop}  =
 \frac{i \tilde{g}^4}{2 N^2 \le \omega_p + \omega_q + p + q + 1 \ri} \,
 \delta_{p+q,r+s} \,  \\
   \left( f^{hbv}  f^{eav}  f^{cew}  f^{dhw}  \, 
  \mathcal{P}_{\ref{fig:1loop_4pt}}  
  +  f^{hbv}  f^{eav}  f^{chw}  f^{dew}  
   \, \mathcal{Q}_{\ref{fig:1loop_4pt}}  \right) \, ,
   \label{eq:I4_result}
\eeq
where
\begin{subequations}
\beq
   \mathcal{P}_{\ref{fig:1loop_4pt}}   =  \left\{
	\begin{aligned}
		&  \frac{2}{p-r}\left[ -h(p) -h(s)+ h(p-r) + h(p-r-1) \right], & p-r \geq 1\\
        & \frac{2}{r-p}\left[ -h(r) -h(q)+ h(r-p) + h(r-p-1) \right], & p-r \leq -1 \\
		&  \sum_{m=1}^{p} \quad \frac{1}{m^2}+\sum_{m=1}^{q}   \frac{1}{m^2} + \le h(p) + h(q) \ri^2 . & p-r=0
	\end{aligned}
	\right.
    \label{eq:I4-line2-result}
\eeq
\beq
   \mathcal{Q}_{\ref{fig:1loop_4pt}}   =  \left\{
	\begin{aligned}
		&  \frac{2}{p-s}\left[-h(p) -h(r)+ h(p-s) + h(p-s-1) \right], & p-s \geq 1\\
        & \frac{2}{s-p}\left[ -h(s) -h(q)+ h(s-p) + h(s-p-1) \right], & p-s \leq -1 \\
		&  \sum_{m=1}^{p} \quad \frac{1}{m^2}+\sum_{m=1}^{q}   \frac{1}{m^2} + \le h(p) + h(q) \ri^2 . & p-s=0
	\end{aligned}
	\right.
    \label{eq:I4-line2-resul2-t}
\eeq
\end{subequations}
We recall that $(p,q)$ are the incoming momenta and $(r,s)$ the outcoming momenta of the fields.
Remarkably, we find that the one-loop correction to the quartic vertex is finite.

\subsection{Higher-loop corrections}
\label{ssec:higher_loops_selfenergy}

Let us consider the higher-loop corrections of the theory.
As discussed in subsection~\ref{ssec:renormalizability_theory}, the counting of the superficial degree of divergence guarantees us that no renormalization is needed when the external lines satisfy $E \geq 4$.
The only non-trivial case is the self-energy, which we analyze below.

Since the theory only admits quartic vertices (see the Feynman rules in subsection~\ref{ssec:Feynman_rules}), we find that the prototypical non-trivial diagram contributing to the quantum correction of the self-energy at loop order $L$ has the form depicted in fig.~\ref{fig:Lloop_selfenergy_scalar}.

\tikzset{every picture/.style={line width=0.75pt}}
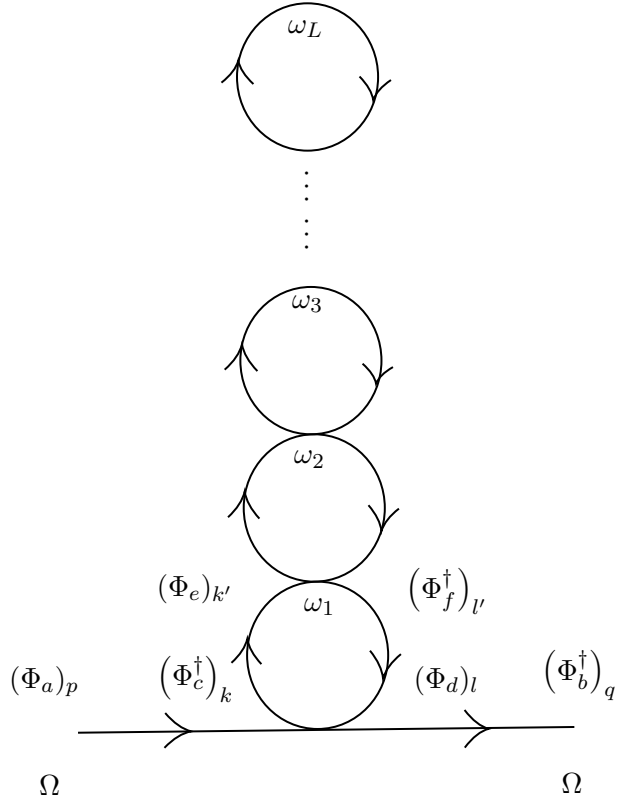
\begin{figure}[ht]
    \centering
    \begin{tikzpicture}[x=0.75pt,y=0.75pt,yscale=-1,xscale=1]

\draw   (225.82,390.59) .. controls (225.82,370.17) and (241.66,353.63) .. (261.21,353.63) .. controls (280.76,353.63) and (296.6,370.17) .. (296.6,390.59) .. controls (296.6,411) and (280.76,427.54) .. (261.21,427.54) .. controls (241.66,427.54) and (225.82,411) .. (225.82,390.59) -- cycle ;
\draw   (303,390) .. controls (298.28,393.87) and (295.29,397.95) .. (294.02,402.21) .. controls (293.56,397.75) and (291.39,393.1) .. (287.48,388.26) ;
\draw   (217.97,394.35) .. controls (222.3,390.4) and (224.95,386.39) .. (225.93,382.3) .. controls (226.62,386.47) and (228.97,390.69) .. (233,395) ;
\draw   (224.13,316.67) .. controls (224.13,296.26) and (239.98,279.71) .. (259.52,279.71) .. controls (279.07,279.71) and (294.91,296.26) .. (294.91,316.67) .. controls (294.91,337.08) and (279.07,353.63) .. (259.52,353.63) .. controls (239.98,353.63) and (224.13,337.08) .. (224.13,316.67) -- cycle ;
\draw   (302,317) .. controls (297.38,320.57) and (294.45,324.33) .. (293.22,328.28) .. controls (292.75,324.14) and (290.58,319.81) .. (286.71,315.28) ;
\draw   (216.25,321.32) .. controls (220.79,317.08) and (223.57,312.78) .. (224.61,308.4) .. controls (225.32,312.86) and (227.79,317.39) .. (232,322) ;
\draw   (222.45,242.75) .. controls (222.45,222.34) and (238.29,205.79) .. (257.84,205.79) .. controls (277.38,205.79) and (293.23,222.34) .. (293.23,242.75) .. controls (293.23,263.16) and (277.38,279.71) .. (257.84,279.71) .. controls (238.29,279.71) and (222.45,263.16) .. (222.45,242.75) -- cycle ;
\draw   (299,246) .. controls (294.48,248.59) and (291.65,251.38) .. (290.52,254.36) .. controls (289.96,251.19) and (287.7,247.83) .. (283.75,244.29) ;
\draw   (214.57,247.29) .. controls (219.3,243.1) and (222.19,238.84) .. (223.26,234.5) .. controls (224.01,238.92) and (226.59,243.42) .. (231,248) ;
\draw   (220.76,100.46) .. controls (220.76,80.05) and (236.61,63.5) .. (256.15,63.5) .. controls (275.7,63.5) and (291.54,80.05) .. (291.54,100.46) .. controls (291.54,120.87) and (275.7,137.42) .. (256.15,137.42) .. controls (236.61,137.42) and (220.76,120.87) .. (220.76,100.46) -- cycle ;
\draw   (298,102) .. controls (293.28,105.17) and (290.31,108.53) .. (289.09,112.1) .. controls (288.55,108.34) and (286.27,104.39) .. (282.23,100.23) ;
\draw   (212.95,103.3) .. controls (217.54,99.68) and (220.36,95.98) .. (221.39,92.2) .. controls (222.15,96.06) and (224.68,99.99) .. (229,104) ;
\draw    (141,429.41) -- (390,426.54) ;
\draw   (334.92,418.01) .. controls (338.97,423.25) and (343.02,426.4) .. (347.06,427.44) .. controls (343.04,428.5) and (339.03,431.64) .. (335.04,436.89) ;
\draw   (184.74,419.53) .. controls (189.05,424.54) and (193.41,427.61) .. (197.81,428.74) .. controls (193.37,429.55) and (188.88,432.3) .. (184.35,437) ;

\draw (120,449.4) node [anchor=north west][inner sep=0.75pt]    {$\Omega$};
\draw (382,448.4) node [anchor=north west][inner sep=0.75pt]    {$\Omega$};
\draw (105,394.4) node [anchor=north west][inner sep=0.75pt]    {$(\Phi _{a})_{p}$};
\draw (372,386.4) node [anchor=north west][inner sep=0.75pt]    {$\left( \Phi _{b}^{\dagger }\right)_{q}$};
\draw (179,389.4) node [anchor=north west][inner sep=0.75pt]    {$\left( \Phi _{c}^{\dagger }\right)_{k}$};
\draw (304,345.4) node [anchor=north west][inner sep=0.75pt]    {$\left( \Phi _{f}^{\dagger }\right)_{l'}$};
\draw (179,348.4) node [anchor=north west][inner sep=0.75pt]    {$( \Phi _{e})_{k'}$};
\draw (308,394.4) node [anchor=north west][inner sep=0.75pt]    {$( \Phi _{d})_{l}$};
\draw (258,145) node [anchor=north west][inner sep=0.75pt]  [rotate=-90]  {$\dotsc \ \dotsc $};
\draw (253,360.4) node [anchor=north west][inner sep=0.75pt]    {$\omega _{1}$};
\draw (248,286.4) node [anchor=north west][inner sep=0.75pt]    {$\omega _{2}$};
\draw (247,208.4) node [anchor=north west][inner sep=0.75pt]    {$\omega _{3}$};
\draw (245,70.4) node [anchor=north west][inner sep=0.75pt]    {$\omega _{L}$};
\end{tikzpicture}
    \caption{Feynman diagram representing the correction at $L$ loops to the self-energy of the scalar field.}
    \label{fig:Lloop_selfenergy_scalar}
\end{figure}

The contribution from the above diagram reads
\beq
\begin{aligned}
 \mathcal{I}^{(2)}_{\ref{fig:Lloop_selfenergy_scalar}} & = 
    \sum_{m_1 \dots m_L =1}^{\infty} \, \sum_{p,p',q,q', \dots=0}^{\infty} 
    \le -\frac{i \tilde{g}^2}{2 N m_1} \ri
    \left[ f^{cav} f^{bdv}  \le  \delta_{p,k+m_1}  \delta_{q,l+m_1} 
    + \delta_{k,p+m_1}  \delta_{l,q+m_1}  \ri  \right. \\
    & \left. 
  -  f^{cdw} f^{abw}  \le  \delta_{p,q+m_1}  \delta_{k,l+m_1} 
    + \delta_{q,p+m_1}  \delta_{l,k+m_1}  \ri 
    \right] \left[ \dots  \right]  \\
& \times    \int_{-\infty}^{\infty} \frac{d \omega_1}{2 \pi}
    \frac{-i \delta_{p,p'} \delta_{ce}}{\omega_1 + p + \frac{1}{2} + i \varepsilon} 
    \frac{-i \delta_{q,q'} \delta_{df}}{\omega_1 + q + \frac{1}{2} + i \varepsilon}  \int_{-\infty}^{\infty} \frac{d\omega_2}{2 \pi}  \, \left( \dots \right) \int_{\infty}^{\infty} \frac{d\omega_L}{2 \pi} \, \left( \dots \right) \, .
\end{aligned}
\label{eq:I2_self_energy_loopL}
\eeq
Importantly, the variable $\omega_1$ only enters the terms written explicitly in eq.~\eqref{eq:I2_self_energy_loopL}, at any loop $L$.
The factors denoted with $(\dots)$ contain additional color structure and propagators inside the other loops, but are independent of $\omega_1$.
Notice that each quartic vertices in the above diagram can belong to one of the two classes depicted in figs.~\eqref{eq:FR_vertexV4} or \eqref{eq:FR_vertexV4tilde}, in any combination.
The following argument is independent of these choices.

Given the structure in eq.~\eqref{eq:I2_self_energy_loopL}, we can explicitly solve the integral over $\omega_1$. The poles of this integral are located on the same side of the $\omega_1$--complex plane:
\beq
\omega_1^{(1)} = - p - \frac{1}{2} - i \varepsilon \, , \qquad
\omega_1^{(2)} = -q-\frac{1}{2} - i \varepsilon \, .
\eeq
Since the integrand scales as $(\omega_1)^{-2}$ at large $\omega_1$, we can apply Jordan's lemma to close the integration contour in the upper half of the complex plane, and then conclude that it vanishes by applying the residue's theorem.
In summary:
\beq
 \int_{-\infty}^{\infty} \frac{d \omega_1}{2 \pi}
    \frac{1}{\omega_1 + p + \frac{1}{2} + i \varepsilon} 
    \frac{1}{\omega_1 + q + \frac{1}{2} + i \varepsilon} = 0 \, \quad \Rightarrow \quad 
    \mathcal{I}_{\ref{fig:Lloop_selfenergy_scalar}}^{(2)} = 0 \, .
\eeq
The above argument applies at any loop order $L \geq 2$, in fact all the integrals over $\omega_1, \omega_2, \dots \omega_{L-1}$ vanish for the same reason.

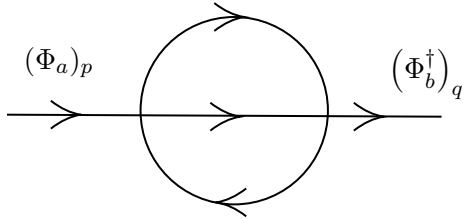
\begin{figure}[ht]
    \centering
    \begin{tikzpicture}[x=0.75pt,y=0.75pt,yscale=-1,xscale=1]
\draw   (171,148) .. controls (171,122.04) and (192.04,101) .. (218,101) .. controls (243.96,101) and (265,122.04) .. (265,148) .. controls (265,173.96) and (243.96,195) .. (218,195) .. controls (192.04,195) and (171,173.96) .. (171,148) -- cycle ;
\draw    (104,150) -- (322,151) ;
\draw   (126,143) .. controls (131,146.89) and (136,149.22) .. (141,150) .. controls (136,150.78) and (131,153.11) .. (126,157) ;
\draw   (206,144) .. controls (211,147.89) and (216,150.22) .. (221,151) .. controls (216,151.78) and (211,154.11) .. (206,158) ;
\draw   (278,144) .. controls (283,147.89) and (288,150.22) .. (293,151) .. controls (288,151.78) and (283,154.11) .. (278,158) ;
\draw   (208,94) .. controls (213,97.89) and (218,100.22) .. (223,101) .. controls (218,101.78) and (213,104.11) .. (208,108) ;
\draw   (224,201) .. controls (219,197.11) and (214,194.78) .. (209,194) .. controls (214,193.22) and (219,190.89) .. (224,187) ;

\draw (111,114.4) node [anchor=north west][inner sep=0.75pt]    {$( \Phi _{a})_{p}$};
\draw (294,115.4) node [anchor=north west][inner sep=0.75pt]    {$\left( \Phi _{b}^{\dagger }\right)_{q}$};

\end{tikzpicture}
    \caption{Additional set of Feynman diagram representing the correction at $L$ loops to the self-energy of the scalar field.}
    \label{fig:Lloop_selfenergy_scalar_2}
\end{figure}

In principle, there is another class of Feynman diagram that we can build with the Feynman rules at our disposal, as depicted in fig.~\ref{fig:Lloop_selfenergy_scalar_2}.
However, similar arguments based on the structure of the poles in the $\omega$ complex plane and the application of residue theorem imply that the corresponding integrals vanish.
The location of the poles of the integrand on the same side of the complex plane is pictorially represented by the fact that arrows (representing the $\mathrm{U}(1)$ charge) in the diagram form a closed loop.

Combining all the above arguments, we have shown that the following selection rule holds:
\begin{srule}
\label{sel_rule1}
Any 1PI-irreducible Feynman diagram at loop order $L \geq 2$ contributing to the quantum corrections to the scalar self-energy vanish identically.  
\end{srule}
In other words, the self-energy does not renormalize!
Similar results about the cancellations of several Feynman diagrams in the context of renormalization of Schr\"{o}dinger-invariant QFTs were found in Refs.~\cite{Bergman:1991hf,Klose:2006dd,Auzzi:2019kdd,Arav:2019tqm,Chapman:2020vtn,Baiguera:2022cbp}.

\begin{figure}[ht]
    \centering     
\begin{tikzpicture}[x=0.75pt,y=0.75pt,yscale=-0.7,xscale=0.7]

\draw   (84,149) .. controls (84,123.04) and (105.04,102) .. (131,102) .. controls (156.96,102) and (178,123.04) .. (178,149) .. controls (178,174.96) and (156.96,196) .. (131,196) .. controls (105.04,196) and (84,174.96) .. (84,149) -- cycle ;
\draw   (121,95) .. controls (126,98.89) and (131,101.22) .. (136,102) .. controls (131,102.78) and (126,105.11) .. (121,109) ;
\draw   (120.06,188.94) .. controls (125.03,192.87) and (130.01,195.24) .. (135,196.06) .. controls (129.99,196.8) and (124.97,199.09) .. (119.94,202.94) ;
\draw   (178,150) .. controls (178,124.04) and (199.04,103) .. (225,103) .. controls (250.96,103) and (272,124.04) .. (272,150) .. controls (272,175.96) and (250.96,197) .. (225,197) .. controls (199.04,197) and (178,175.96) .. (178,150) -- cycle ;
\draw   (215,96) .. controls (220,99.89) and (225,102.22) .. (230,103) .. controls (225,103.78) and (220,106.11) .. (215,110) ;
\draw   (214.06,189.94) .. controls (219.03,193.87) and (224.01,196.24) .. (229,197.06) .. controls (223.99,197.8) and (218.97,200.09) .. (213.94,203.94) ;
\draw   (357,148) .. controls (357,122.04) and (378.04,101) .. (404,101) .. controls (429.96,101) and (451,122.04) .. (451,148) .. controls (451,173.96) and (429.96,195) .. (404,195) .. controls (378.04,195) and (357,173.96) .. (357,148) -- cycle ;
\draw   (394,94) .. controls (399,97.89) and (404,100.22) .. (409,101) .. controls (404,101.78) and (399,104.11) .. (394,108) ;
\draw   (393.06,187.94) .. controls (398.03,191.87) and (403.01,194.24) .. (408,195.06) .. controls (402.99,195.8) and (397.97,198.09) .. (392.94,201.94) ;
\draw    (19,88) -- (84,149) ;
\draw    (451,148) -- (516,209) ;
\draw    (451,148) -- (517,100) ;
\draw    (18,197) -- (84,149) ;

\draw (285,136.4) node [anchor=north west][inner sep=0.75pt]    {$\dotsc \ \dotsc $};
\draw (35,64.4) node [anchor=north west][inner sep=0.75pt]    {$( \Phi _{a})_{p}$};
\draw (36,203.4) node [anchor=north west][inner sep=0.75pt]    {$( \Phi _{b})_{q}$};
\draw (468,64.4) node [anchor=north west][inner sep=0.75pt]    {$\left( \Phi _{c}^{\dagger }\right)_{r}$};
\draw (462,195.4) node [anchor=north west][inner sep=0.75pt]    {$\left( \Phi _{d}^{\dagger }\right)_{s}$};
\end{tikzpicture}
    \caption{Feynman diagram contributing to the quantum correction at $L$ loops to the quartic vertex of the scalar field.}
    \label{fig:Lloop_4pt_scalar}
\end{figure}
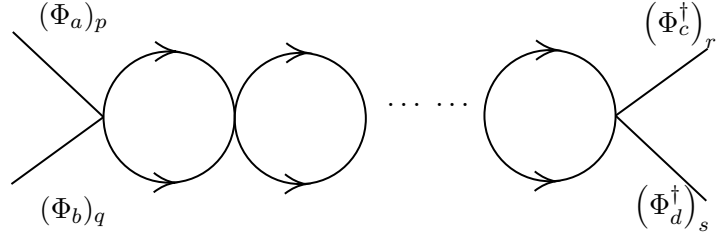

We briefly comment on the structure of higher-loop corrections to the quartic vertex.
By the above arguments, the only non-vanishing contributions at loop order $L$ arise from the class depicted in fig.~\ref{fig:Lloop_4pt_scalar}.
In this set of diagrams, the vertices can be taken in any combination among the two structure in eqs.~\eqref{eq:FR_vertexV4} or \eqref{eq:FR_vertexV4tilde}.
Specializing the arguments of subsection~\ref{ssec:renormalizability_theory} to $E=4$ and $L \geq 2$, one can show that these diagrams always give a finite result.

This concludes the study of perturbative quantum corrections to the SU(1,1) bosonic SMT using the Feynman diagrammatic approach.
We summarize and discuss the results in section~\ref{sec:conclusions}.

\section{Quantum corrections of SU(1,1) SMT (QM approach)}
\label{sec:quantum_SMT_QM}

In section~\ref{sec:quantum_SU11_QFT}, we computed the perturbative quantum corrections to the bosonic SU(1,1) sector using a diagrammatic approach and working with the path integral in terms of classical fields, see discussion around eq.~\eqref{eq:effective_action}.
In the following, we will match the previous results with an alternative method to compute loop corrections: by treating fields as quantum operators and employing time-dependent perturbation theory.
We briefly review this method in section~\ref{ssec:time_dep_perturbation}, and then employ it to compute scattering amplitudes in sections~\ref{ssec:check_11_scattering} ($1 \rightarrow 1$ particles) and \ref{ssec:check_22_scattering} ($2 \rightarrow 2$ particles).

\subsection{Time-dependent perturbation theory}
\label{ssec:time_dep_perturbation}

To compare the two approaches -- where fields are treated as classical variables or quantum operators -- we evaluate scattering amplitudes involving $1 \rightarrow 1$ particles (self-energy) or $2 \rightarrow 2$ particles (quartic vertex).
In standard QFT, scattering experiments are studied via wave packets prepared at asymptotically early and late times, leading to the LSZ reduction formula~\cite{Peskin:1995ev,Srednicki:2007qs,Schwartz:2014sze}.
By appropriately adapting the procedure to the SU(1,1) bosonic SMT, we find that the scattering amplitude for a process involving $M$ initial particles and $N$ final particles reads
\begin{equation}
\begin{aligned}
& \langle f_1 \cdots f_N |S| i_1 \cdots i_M  \rangle = \\
& = \left[ i \int dt_{f_1} e^{-i\omega_{f_1} t_{f_1}} \left( i\partial_{t_{f_1}}+n_1+\frac{1}{2} \right)  \right] \cdots 
\left[ i \int dt_{f_N} e^{-i\omega_{f_N} t_{f_N}} \left( i\partial_{t_{f_N}}+n_N+\frac{1}{2} \right)  \right]  \\
& \times \left[ i \int dt_{i_1} e^{i\omega_{i_1} t_{i_1}} \left( i\partial_{t_{i_1}}+m_1+\frac{1}{2} \right)  \right] \cdots 
\left[ i \int dt_{i_M} e^{i\omega_{i_M} t_{i_M}} \left( i\partial_{t_{i_M}}+m_M+\frac{1}{2} \right)  \right] 
\\
& \times \langle \Omega | T \{ a_{n_1} (t_{f_1}) \cdots a_{n_{N}} (t_{f_N}) a^{\dagger}_{m_1} (t_{i_1}) \cdots a^{\dagger}_{m_M} (t_{i_M})   \} | \Omega \rangle  \, ,
\end{aligned}
\label{eq:LSZ_formula_SMT}
\end{equation}
where we denoted with $S$ the S--matrix, $\ket{\Omega}$ the interacting vacuum of the theory, and $T$ the time-ordering.
Notice that here we consider the SU(1,1) SMT after the expansion over the Fourier modes associated with the compact spatial direction, in other words the fields~\eqref{Phi_field} only depend on the temporal coordinate.
Each square parenthesis contains a factor proportional to the on-shell kinetic operator.

Next, we can express the previous formula in terms of expectation values on the free vacuum $\ket{0}$ as follows:
\beq
\begin{aligned}
& \langle\Omega| T \{ a_{n_1} (t_{f_1}) \cdots a_{n_{N}} (t_{f_N}) a^{\dagger}_{m_1} (t_{i_1}) \cdots a^{\dagger}_{m_M} (t_{i_M})   \} |\Omega\rangle = \\
& = \frac{\langle 0|T\left\{ a_{n_1} (t_{f_1}) \cdots a_{n_{N}} (t_{f_N}) \exp [-i \int_{-\infty}^{\infty} d t (H_{\rm int})_I(t)]  a^{\dagger}_{m_1} (t_{i_1}) \cdots a^{\dagger}_{m_M} (t_{i_M})   \right\}|0\rangle}{\langle 0|T\left\{\exp [-i \int_{-\infty}^{\infty} d t (H_{\rm int})_I(t)] \right\}|0\rangle} \, .
\end{aligned}
\eeq
In the above expression, the subscript $I$ refers to the \textit{interaction picture}. An operator $\mathcal{O}_I$ in the interaction picture is defined in terms of the Schr\"{o}dinger picture by
\beq
\mathcal{O}_{\rm I} (t) \equiv e^{i H_0 t} \mathcal{O} e^{-i H_0 t} \, ,
\eeq
where $H_0$ is the free Hamiltonian.

In this context, SMTs play a special role. Since they are effective theories obtained as near-BPS limits satisfying the conditions~\eqref{eq:SMT_dilatation_operator}, by virtue of the commutators~\eqref{eq:commutators_dilatation} we have
\beq
[H_0, H_{\rm int}] = 0 \, , \quad \, 
(H_{\rm int})_I (t) = H_{\rm int} \, .
\label{eq:simplification_Hint_interacting_picture}
\eeq
In other words, the interacting Hamiltonian in the interaction picture is time-independent.

Let us recall that the kinetic operators in the  LSZ reduction formula~\eqref{eq:LSZ_formula_SMT} cancel the on-shell propagators in the external particle states (\eg see discussion in section~6.1.1 of \cite{Schwartz:2014sze}).
Combining this observation with the result~\eqref{eq:simplification_Hint_interacting_picture}, we obtain
\begin{equation}
\begin{aligned}
& \langle f_1 \cdots f_N |S| i_1 \cdots i_M  \rangle = \\
& = \left[ i \int dt_{f_1} e^{-i\omega_{f_1} t_{f_1}}   \right] \cdots 
\left[ i \int dt_{f_N} e^{-i\omega_{f_N} t_{f_N}}  \right]   \left[ i \int dt_{i_1} e^{i\omega_{i_1} t_{i_1}} \right] \cdots 
\left[ i \int dt_{i_M} e^{i\omega_{i_M} t_{i_M}}  \right] 
\\
& \times \frac{\langle 0|T\left\{ a_{n_1}  (t_{f_1}) \cdots a_{n_{N}}  (t_{f_N})\exp [-i \int_{-\infty}^{\infty} d t \, H_{\rm int} ]  a^{\dagger}_{m_1}  (t_{i_1}) \cdots a^{\dagger}_{m_M}   (t_{i_M})  \right\}|0\rangle}{\langle 0|T\left\{\exp [-i \int_{-\infty}^{\infty} d t \, H_{\rm int}] \right\}|0\rangle} \, .
\end{aligned}
\label{eq:LSZ_formula_SMT2}
\end{equation}
Our goal is to apply perturbation theory to compute the quantum corrections to the effective action using this QM formalism.

\subsection{QM computation of $1 \rightarrow 1$ scattering process}
\label{ssec:check_11_scattering}

In order to extract the quantum corrections to the self-energy, we need to consider a $1 \rightarrow 1$ scattering process.
To extract the one-loop contribution, we consider the first order in the perturbative expansion of the exponential in eq.~\eqref{eq:LSZ_formula_SMT2}, \ie
\beq
 \langle f | S  | i \rangle  =
 i \left[ \int dt_1 e^{-i \omega_{q} t_1}  \right] \left[ \int dt_2 e^{i\omega_{p} t_2}\right] \times \langle 0 | T \left\lbrace a_{q} (t_1)
 \left(  \int_{-\infty}^{\infty} d t_3 \, H_{\text{int}}  \right)
 a_p^\dagger (t_2) \right\rbrace | 0 \rangle \,. 
 \label{eq:LSZ_11_particles}
\eeq
Using the first-order term in perturbation theory is consistent with the fact that the one-loop diagrams in section~\ref{ssec:1loop_selfenergy} always contain one quartic vertex.
The expression~\eqref{eq:LSZ_11_particles} is proportional to
\beq
\langle 0 |  a_{q}  H_{\rm int}  a^{\dagger}_p   |  0 \rangle = 0 \, , 
\eeq
where we used the fact that $H_{\rm int}$ is normal-ordered, of the form $H_{\rm int} \sim a^{\dagger} a^{\dagger} a a$. 
Even though we can get rid of one annihilation operator by acting on the one-particle state via the commutation relation~\eqref{eq:commutation_relations_ladder_operators_SMT}, there is always one surviving lowering operator acting on the vacuum, which vanishes.

At higher loops, we need to consider higher orders in perturbation theory.
Since at loop order $L$ the resulting expressions are proportional to
\beq
\langle 0 |  a_{q}  (H_{\rm int})^L  a^{\dagger}_p   |  0 \rangle = 0 \, , 
\eeq
we find again a vanishing result by normal ordering. Therefore, we conclude that there are no quantum corrections to the self-energy.
This result is consistent with the computations performed in sections~\ref{ssec:1loop_selfenergy} and \ref{ssec:higher_loops_selfenergy}.

\subsection{QM computation of $2 \rightarrow 2$ scattering process}
\label{ssec:check_22_scattering}

To extract the quantum corrections to the quartic vertex, we need to consider a $2 \rightarrow 2$ scattering process.
In particular, the one-loop contribution corresponds to the second order in perturbation theory, as confirmed by the appearance of two vertices 
in the Feynman diagrams in section~\ref{ssec:1loop_quartic_vertex}.
Concretely, we need to compute
\begin{align}
& \langle f_1 f_2 | S| i_1 i_2 \rangle = - \frac{1}{2} \int dt_1 e^{-i \omega_{r} t_1} \int dt_2 e^{-i\omega_{s} t_2}  \int dt_3 e^{i\omega_{p} t_3} \int dt_4 e^{i\omega_{q} t_4} \nonumber \\
&\times \langle 0 | T \left\lbrace a_{r}(t_1) a_{s}(t_2) 
\left( \int_{-\infty}^{\infty} d t_5 \, H_{\text{int}} \int_{-\infty}^{\infty} d t_6 \, H_{\text{int}}  \right)
a_p^\dagger(t_3) a_q^\dagger(t_4) \right\rbrace | 0 \rangle \, .
\label{eq:4pt_amplitude_1loop_QM}
\end{align}
In the following, we are interested in extracting a contribution to the effective action of the form~\eqref{eq:def_quartic_vertex_from_effective_action}, where the color structure of the external fields is factored out.
Similar to their classical counterpart, quantum operators belong to the adjoint representation of $\mathrm{SU}(N)$ and are decomposed as $a^{i}_{\,\,j} \equiv a^a (T_a)^i_{\,\,j}$.

We recall that the interacting Hamiltonian can be written in the general form~\eqref{eq:dictionary_bos_term_SMT}, where the coefficients $U^{s'r'}_{sr}$ in the bosonic SU(1,1) sector read
\beq
\begin{aligned}
    U^{mn}_{mn} & =-\frac{1}{2}(h(m)+h(n))  \, , \\
    U^{m-l,n+l}_{mn} & =\frac{1}{2l} \quad \text{ for }l=1,\cdots,m\\
    U^{m+l,n-l}_{mn} & =\frac{1}{2l} \quad \text{ for }l=1,\cdots,n
\end{aligned}
\label{eq:dictionary_U_su11}
\eeq
The interacting Hamiltonian acts on two-particle states as follows:
\beq
H_{\rm int} {(a^\dagger_{p})}_{a} {(a^\dagger_{q})}_{b}  | 0 \rangle = -\frac{1}{2N} \left( U_{pq}^{ij} f^{aev} f^{efv}+ U_{qp}^{ij}f^{cev} f^{fav} \right) (a_{s'}^\dagger)_e (a_{r'}^\dagger)_f|0\rangle \, ,
\label{eq:Hint_on_2particles}
\eeq
where we used eq.~(2.18) of Ref.~\cite{Harmark:2014mpa}, the group-theoretical identity $[T_a, T_b] = i f^{abc} T_c$, and the normalization of the trace in eq.~\eqref{eq:fields_matrices}.
The antisymmetry of the structure constants and the normal ordering in the Hamiltonian have already been employed to simplify the above result.

Plugging the identity~\eqref{eq:Hint_on_2particles} inside eq.~\eqref{eq:4pt_amplitude_1loop_QM}, we find
\beq
\begin{aligned}
    \langle f_1 f_2 | S| i_1 i_2 \rangle & =
    \langle 0| {(a_{r})}_{c} {(a_{s})}_{d} \, H_{\text{int}}^2 \, {(a^\dagger_{p})}_{a} {(a^\dagger_{q})}_{b}  | 0 \rangle = \\
    & = \frac{1}{N^2} \left( U_{rs}^{ij} U_{pq}^{ij} f^{cuv} f^{ydv}  f^{auw} f^{ybw}+U_{sr}^{ij} U_{pq}^{ij}f^{cyv}f^{udv}f^{auw}f^{ybw} \right) \, .
\end{aligned}
\label{eq:step1_1loop_quartic_QM}
\eeq
Next, we need to sum over all the possible positive integers $i$ and $j$ for given external momentum $p$, $q$, $r$ and $s$. 
This summation is performed case by case in appendix~\ref{app:ssec:QM_computation_22}, finding the following result:
\beq
\langle f_1 f_2 | H_{\text{int}}^2| i_1 i_2 \rangle=\frac{\delta_{p+q,r+s}}{2N^2} \left( \mathcal{P}_{\ref{fig:1loop_4pt}} f^{gbv} f^{chv}  f^{ebw} f^{cfw}+\mathcal{Q}_{\ref{fig:1loop_4pt}}f^{gcv}f^{bhv}f^{ebw}f^{cfw} \right)  \, ,
\label{eq:step2_1loop_quartic_QM}
\eeq
where $\mathcal{P}_{\ref{fig:1loop_4pt}}$ and $\mathcal{Q}_{\ref{fig:1loop_4pt}}$ were defined in eqs.~\eqref{eq:I4-line2-result} and \eqref{eq:I4-line2-resul2-t}.

Therefore, the scattering amplitude is given by
\beq
\begin{aligned}
\langle r s | S | p q  \rangle_{\rm 1-loop} & = 
- \frac{1}{2} \int_{-\infty}^{\infty} d t_5\int_{-\infty}^{t_5} d t_6 \times e^{-i(r+s+1)t_5} e^{i(p+q+1)t_6} \langle f_1 f_2|H_{\text{int}}^2|i_1 i_2 \rangle = \\
& =  \frac{i \, \delta_{p+q,r+s}}{2(p+q+1)}\langle f_1 f_2|H_{\text{int}}^2|i_1 i_2 \rangle = \\
& =  \frac{i}{4N^2} \frac{\delta_{p+q,r+s}}{p+q+1} (\mathcal{P}_{\ref{fig:1loop_4pt}} f^{gbv} f^{chv}  f^{ebw} f^{cfw}+\mathcal{Q}_{\ref{fig:1loop_4pt}}f^{gcv}f^{bhv}f^{ebw}f^{cfw})
\, .
\end{aligned}
\eeq
This result matches with eq.~\eqref{eq:I4_result}, after putting the external momenta on-shell!

One can repeat the same procedure in the case of higher-loop corrections to the quartic vertex.
This computation would require to consider higher-order terms in the series-expansion of the amplitude~\eqref{eq:LSZ_formula_SMT2}.
Here, we simply point out that the analysis of the superficial degree of divergence in section~\ref{ssec:renormalizability_theory} tells us that no UV divergences occur.
Nevertheless, we checked in some explicit cases that the matching between this formalism and the diagrammatic approach in section~\ref{sec:quantum_SU11_QFT} continues to hold at higher loops.

\section{Conclusions}
\label{sec:conclusions}

In this paper, we investigated the classical and quantum properties of SU(1,1) bosonic SMT, a non-Lorentzian model obtained from a near-BPS decoupling limit of $\mathcal{N}=4$ SYM.

Firstly, we determined an action formulation of the theory in terms of two-dimensional \text{semi-local} fields on $\mathbb{R} \times S^1$.
We have shown that the classical action is off-shell invariant under the SU(1,1) group, found the corresponding transformations of the coordinates and fields, and computed the associated Noether charges.

Secondly, we studied the renormalization properties of the SU(1,1) bosonic SMT after expanding the fields in Fourier modes over the spatial direction $S^1$.
We performed the computation using two methods: a Feynman diagrammatic approach in terms of commuting fields within the path integral, and time-dependent perturbation theory in terms of quantum-mechanical operators.
We have shown that the outcomes obtained by using the two methods match.
The results for the quantum corrections of the effective action are summarized in table~\ref{tab:summary_corrections}.

\begin{table}[ht]   
\begin{center}   
\begin{tabular}  {|p{42mm}|c|c|c|} \hline  
 & \textbf{Self-energy}  & \textbf{4-point vertex}   \\ \hline \rule{0pt}{4.9ex}
\rule{0pt}{4.9ex} 1--loop corrections  &  \makecell{0 \\
(section~\ref{ssec:1loop_selfenergy})} &  \, \makecell{Finite \\ (section~\ref{ssec:1loop_quartic_vertex}) }  \\
\rule{0pt}{4.9ex} Higher--loop corrections   &  \, \makecell{0 \\ (section~\ref{ssec:higher_loops_selfenergy}) } & \makecell{Finite \\ 
(section~\ref{ssec:higher_loops_selfenergy})}  \\[0.2cm]
\hline
\end{tabular}   
\caption{Summary of the quantum corrections to the bosonic SU(1,1) Spin Matrix Theory.} 
\label{tab:summary_corrections}
\end{center}
\end{table}

We found that the loop corrections to the self-energy vanish at all orders in perturbation theory.
From the quantum-mechanical perspective, this result is a consequence of the fact that the interacting Hamiltonian of the theory is normal-ordered.
From the Feynman diagrammatic perspective, it is a consequence of the causal structure of the non-Lorentzian propagator, together with the $\mathrm{U}(1)$ particle number conservation.

Furthermore, we found that the quartic vertex only admits finite quantum corrections at any order in perturbation theory.
Since the quantum corrections are a function of the external momenta and no renormalization scale needs to be introduced, we conclude that the beta function associated with the coupling constant of the theory vanishes ($\beta_{\tilde{g}}=0$).
In other words, the theory is perturbatively \textit{finite} and a non-renormalization theorem holds.

Non-renormalization theorems often arise within Schr\"{o}dinger-invariant QFTs. 
For instance, the self-energy does not renormalize in the three-dimensional theories studied in Refs.~\cite{Bergman:1991hf,Klose:2006dd}, which contain a complex scalar field interacting via a quartic potential.
The vertex of the theory receives UV-divergent contributions at all orders in perturbation theory, which can be resummed.
When including $\mathcal{N}=2$ supersymmetry, one can build a Galilean analog of the Wess-Zumino model~\cite{Auzzi:2019kdd}.
This theory has a one-loop exact self-energy, while the superpotential does not renormalize.
Similar non-renormalization theorems have been found in the non-Lorentzian settings studied in Refs.~\cite{Arav:2019tqm,Chapman:2020vtn,Baiguera:2022cbp}.
In all these cases, either the self-energy or the vertices receive certain UV-divergent loop corrections.

On the contrary, SU(1,1) bosonic Spin Matrix Theory provides a rare example of a non-supersymmetric and non-Lorentzian theory that is finite at all orders in perturbation theory.
This result is consistent with the fact that the effective Hamiltonian only contains contributions from the tree-level and the one-loop dilatation operator of the $\mathcal{N}=4$ super Yang-Mills parent theory~\cite{Harmark:2007px}.
All the (finite) quantum corrections of SU(1,1) Spin Matrix Theory arise from an exponentiation of the one-loop exact effective Hamiltonian, and there are no novel UV divergences.

\subsection*{Outlook}

We discuss various possible extensions and implications of our work:
\begin{enumerate}
    \item \textbf{Extension to other Spin Matrix Theories.}
    In this work, we investigated the simplest example of SMT having the degrees of freedom compatible with a $(1+1)$--dimensional QFT.
    Building on these methods, we plan to study the classical symmetries and the renormalization properties of richer SMT limits.
    The most natural follow-ups are the SU(1,1) fermionic sector and the  SU$(1,1|1)$ sector, the latter being the simplest supersymmetric case.
    It would be interesting to determine the intertwining between the non-renormalization properties of SMT and supersymmetry. A superspace formulation for the SU$(1,1|1)$ SMT was proposed in Ref.~\cite{Baiguera:2020jgy}.

    Next, the challenge would be to consider SMT limits leading to $(2+1)$--dimensional theories, which all share a bosonic $\mathrm{SU}(1,2)$ subgroup.
    The authors of Ref.~\cite{Baiguera:2024vlj} found that a natural background where such QFTs could be defined is given by a null-reduction of $\mathbb{R} \times S^3$, and they also determined a mapping between state and operator pictures.
    Finding a QFT formulation for these sectors is an open problem.
    \item \textbf{Strong coupling.}
    As shown in Refs.~\cite{Baiguera:2020jgy,Baiguera:2020mgk,Baiguera:2021hky,Baiguera:2022pll}, the interacting Hamiltonian of any SMT is positive-definite and can be written as a quadratic expression in terms of certain fundamental blocks.
    In the SU(1,1) bosonic sector, the structure reads
    \beq
    H = H_0 + \frac{\tilde{g}^2}{2N} \sum_{l=1} \frac{1}{l} \tr \le j^{\dagger}_l j_l \ri \, ,
    \label{eq:recap_Hint}
    \eeq
    where we are ignoring terms that vanish on physical states.
    The identity~\eqref{eq:recap_Hint} implies that the leading contribution in the strong-coupling limit $\tilde{g} \rightarrow \infty$ corresponds to set $j_l=0$ for $l \geq 0$.
    It would be interesting to understand whether this identity could simplify the study of the strong-coupling behavior of the theory, and find the corresponding constraints on the semi-local formulation of the theory.
    \item \textbf{Microstate counting.}
    The largest symmetry group of the parent $\mathcal{N}=4$ SYM theory that a SMT limit can preserve is $\mathrm{PSU}(1,2|3)$.
 Via the AdS/CFT correspondence, it is expected that this sector can capture information about the microstates of $1/16$--BPS black hole solutions that exist in five-dimensional AdS geometry~\cite{Gutowski:2004yv}.
 In particular, recent work classified the operators of $\mathcal{N}=4$ in terms of the so-called monotone and fortuitous operators, the latter encoding information about black hole microstates~\cite{Chang:2013fba,Chang:2022mjp,Choi:2022caq,Choi:2023znd,Choi:2023vdm, Chang:2024zqi,Chang:2025mqp} (see also the related work~\cite{Lei:2026fep}).
 We expect that a QFT formulation of the PSU$(1,2|3)$ SMT could allow to better understand this microstate counting, and even study the case of near-BPS solutions.
 \item \textbf{Holographic matching.}
 One of the main advantages of performing a near-BPS limit is that several modes of $\mathcal{N}=4$ SYM decouple, and the resulting SMT is easier to handle. 
 In the SU(1,1) bosonic sector, this hope is nourished by the fact that the theory is finite at quantum level (as we have shown in this work), thus making simpler to move from weak to strong coupling.
 Our next goal is to formulate a more precise holographic correspondence which would match field theory computations with the string theory side. 
 For instance, in the planar limit the bosonic SU(1,1) sector admits a dual $\sigma$-model defined on a $\mathrm{U}(1)$ Galilean geometry given by $\mathbb{R}$ (the time direction) times a cigar-geometry~\cite{Harmark:2017rpg,Harmark:2018cdl,Harmark:2019upf,Harmark:2020vll}. One could hope to extend this link beyond the planar limit.
 Another possibility is to study the emergence of D-branes in the form of dual giant gravitons, extending the studies of Ref.~\cite{Harmark:2016cjq}.  
 Finally, one could compute quantum information probes, such as entanglement or complexity (for some reviews on these topics, \eg see~\cite{Horodecki:2009zz,Rangamani:2016dms,Susskind:2018pmk,Nandy:2024evd,Baiguera:2025dkc,Rabinovici:2025otw}). Concretely, Ref.~\cite{Das:2024tnw} computed the spread complexity -- a measure of the expansion of states across the Hilbert space during time evolution -- for semiclassical strings in $\mathrm{AdS}_5 \times S^5$ in the planar limit.
 The SMT sectors could provide a setting to perform these computations at finite $N$.
\end{enumerate}

\section*{Acknowledgements}
We thank Yang Lei and Ziqi Yan for valuable discussions.
RA, LG and GN are supported in part by
INFN through the \textit{Gauge and String Theory} (GAST) research project.
The work of SB is supported by the INFN grant \textit{Gauge Theories and Strings} (GAST) via a research grant on \textit{Holographic dualities, quantum information and gravity}.
TH acknowledges support from the “Center of Gravity”, a Center
of Excellence funded by the Danish National Research Foundation under Grant N.~DNRF184.

\appendix

\section{Technical tools}
\label{app:math_tools}

This appendix contains additional technical details involving the one-loop correction to the four-point vertex.

\subsection{Details for the one-loop correction to the four-point vertex}
\label{app:ssec:details_1loop_4pt_vertex}

In this section, we provide more technical details for the computation of the one-loop corrections to the quartic vertex,  listed in subsection~\ref{ssec:1loop_quartic_vertex} and pictorially depicted in fig.~\ref{fig:1loop_4pt}.

\paragraph{Diagram in fig.~\ref{subfig:1loop_4pt_old_old}.}
Starting from the contribution in eq.~\eqref{eq:1loop_4pt_contr1}, we plug in the result~\eqref{eq:integral_4pt_residue} of the integration over the $\omega$ space and we perform the products between the structure constants and the Kronecker $\delta$ inside the parentheses.
After tedious but straightforward manipulations, we get
\beq
\begin{aligned}
 & \mathcal{I}^{(4)}_{\ref{subfig:1loop_4pt_old_old}}  =
 \frac{i \tilde{g}^4}{2 N^2 \le \omega_p + \omega_q + p + q + 1 \ri} \,
 \\
& \times 
\left[  f^{hbv}  f^{eav}  f^{cew}  f^{dhw}  
\le  \sum_{m=1}^{p} \sum_{n=1}^{r} 
 \delta_{p+n,r+m}  \delta_{q+m, s+n} 
  + \sum_{m=1}^{p} \sum_{n=1}^{s} 
  \delta_{p-m, r+n} \delta_{q+m, s-n} 
  \right.  \right. \\ 
  & \left. \left. 
  + \sum_{m=1}^{q} \sum_{n=1}^{r}  
  \delta_{p+m,r-n} \delta_{q-m,s+n} 
  + \sum_{m=1}^{q} \sum_{n=1}^{s}
  \delta_{p+m,r+n} \delta_{q+n, s+m} 
  \ri \right. \\
& \left. + f^{hbv}  f^{eav}  f^{chw}  f^{dew} 
\le   \sum_{m=1}^{p} \sum_{n=1}^{r} 
\delta_{p-m, s+n}  \delta_{q+m, r-n}
+ \sum_{m=1}^{p} \sum_{n=1}^{s}
\delta_{p+n, s+m} \delta_{q+m, r+n}  \right. \right. \\
& \left. \left. 
+  \sum_{m=1}^{q} \sum_{n=1}^{r}
\delta_{p+m,s+n} \delta_{q+n, r+m} 
+ \sum_{m=1}^{q} \sum_{n=1}^{s}  
\delta_{p+m, s-n} \delta_{q-m, r+n} 
\ri  \right]_{\rm constr.}  \times \frac{1}{m n}
\, ,
\end{aligned}
\label{eq:I4a_starting2}
\eeq
where the subscript \textit{constr.} reminds us that we need to impose that 
all the integer momenta entering the above expressions are positive, both for physical and virtual particles.

All the above terms contain two factors of Kronecker $\delta$. One of them can be used to perform the summations over the spatial momenta $n$ of the fields inside the loop, leading to an overall factor $\delta_{p+q,r+s}$ of total momentum conservation.
We evaluate the remaining summation over $m$ below.

Let us now analyze the precise ranges of summation of each term in the first square parenthesis of eq.~\eqref{eq:I4a_starting2}:
\begin{itemize}
\item \textbf{Term 1.} We have $n=m-p+r$. Since $1 \leq n \leq r$, we have $m \geq p-r+1$ and $m \leq p$ (a condition already imposed). There are now the following subcases:
    \beq
    \sum_{m=1}^{p} \sum_{n=1}^{r} \delta_{p+n,r+m}  \delta_{q+m, s+n} \, \frac{1}{m n}=
    \delta_{p+q,r+s}
    \left\{
	\begin{aligned}
		& \sum_{m=p-r+1}^{p} \frac{1}{m(m-p+r)}, & p-r>0\\
        & \quad \sum_{m=1}^{p} \quad \frac{1}{m^2} , & p-r=0\\
		& \quad \sum_{m=1}^{p} \quad \frac{1}{m(m-p+r)} , & p-r<0
	\end{aligned}
	\right.
    \label{eq:term1_4ptvertex_1loop_app}
    \eeq
\item \textbf{Term 2.}  We have $n=-m+p-r$. Since $1 \leq n \leq s$, we get $m \leq p-r-1$ and $m \geq -q$ (trivially satisfied). This leads to the following subcases:
    \beq
    \sum_{m=1}^{p} \sum_{n=1}^{s} \delta_{p-m, r+n} \delta_{q+m, s-n} \, \frac{1}{m n}=
    \delta_{p+q,r+s}
    \left\{
	\begin{aligned}
		& \sum_{m=1}^{p-r-1} \frac{1}{m(-m+p-r)}, & p-r>1\\
        & \quad 0, & p-r\leq 1
	\end{aligned}
	\right.
     \label{eq:term2_4ptvertex_1loop_app}
    \eeq
\item \textbf{Term 3.}  We have $n=-m-p+r$. Since $1 \leq n \leq r$, we find $m \leq r-p-1$ and $m \geq -p$ (trivially satisfied). The summation reduces to the following subcases:
    \beq
    \sum_{m=1}^{q} \sum_{n=1}^{r} \delta_{p+m,r-n} \delta_{q-m,s+n} \, \frac{1}{m n}=
     \delta_{p+q,r+s}
    \left\{
	\begin{aligned}
		& \quad 0, & p-r \geq -1\\
        & \sum_{m=1}^{-p+r-1} \frac{1}{m(-m-p+r)}, & p-r\leq -1
	\end{aligned}
	\right.
     \label{eq:term3_4ptvertex_1loop_app}
    \eeq
\item \textbf{Term 4.}  We have $n=m+p-r$. Since $1 \leq n \leq s$, we get $m \geq -p+r+1$ and $m \leq q$ (a condition already imposed). In summary, we find 
    \beq
    \sum_{m=1}^{q} \sum_{n=1}^{s} \delta_{p+m,r+n} \delta_{q+n, s+m} \, \frac{1}{m n}=
     \delta_{p+q,r+s}
    \left\{
	\begin{aligned}
		&  \quad\sum_{m=1}^{q}  \quad \frac{1}{m(m+p-r)}, & p-r>0\\
        & \quad \sum_{m=1}^{q} \quad \frac{1}{m^2} , & p-r=0\\
		& \sum_{m=-p+r+1}^{q} \frac{1}{m(m+p-r)}, & p-r<0
	\end{aligned}
	\right.
     \label{eq:term4_4ptvertex_1loop_app}
    \eeq
\end{itemize}
The terms in the second square parenthesis of eq.~\eqref{eq:I4a_starting2}
can be treated similarly, with the external momentum $p$ replaced by $s$.
Summing the contributions from eqs.~\eqref{eq:term1_4ptvertex_1loop_app}--\eqref{eq:term4_4ptvertex_1loop_app}, we find the following list of cases:
\beq
    \left\{
	\begin{aligned}
		&  \sum_{m=p-r+1}^{p} \frac{1}{m(m-p+r)}+\sum_{m=1}^{p-r-1} \frac{1}{m(-m+p-r)}+\sum_{m=1}^{q} \quad  \frac{1}{m(m+p-r)}, & p-r>1\\
        & \quad \sum_{m=2}^{p} \frac{1}{m(m-1)}+\sum_{m=1}^{q} \quad  \frac{1}{m(m+1)}, & p-r=1\\
		& \quad \sum_{m=1}^{p} \quad \frac{1}{m^2}+\sum_{m=1}^{q} \quad  \frac{1}{m^2}, & p-r=0\\
        & \quad \sum_{m=1}^{p} \quad \frac{1}{m(m+1)}+\sum_{m=-2}^{q} \frac{1}{m(m-1)}, & p-r=-1\\
        &  \quad \sum_{m=1}^{p} \quad \frac{1}{m(m-p+r)} + \sum_{m=1}^{-p+r-1} \frac{1}{m(-m-p+r)} + \sum_{m=-p+r+1}^{q} \frac{1}{m(m+p-r)}. & p-r<-1
	\end{aligned}
	\right.
    \label{I4a-line2}
\eeq
For concreteness, we show how to evaluate the summation in the first line, \ie when $p-r >1$.
By separating the terms in simple fractions and employing the definition of harmonic numbers, we obtain
\beq
\begin{aligned}
& \sum_{m=p-r+1}^{p} \frac{1}{m(m-p+r)}+\sum_{m=1}^{p-r-1} \frac{1}{m(-m+p-r)}+\sum_{m=1}^{q} \quad  \frac{1}{m(m+p-r)} = \\
= & \frac{1}{p-r}\left[ \sum_{m=p-r+1}^{p} \left( \frac{1}{m-p+r}-\frac{1}{m} \right) + \sum_{m=1}^{p-r-1} \left( \frac{1}{m}+\frac{1}{p-r-m} \right)+ \sum_{m=1}^{q} \left( \frac{1}{m}-\frac{1}{m+p-r} \right) \right] = \\
= & \frac{1}{p-r}\left[ h(r)-h(p)+h(q)-h(s)+ 2h(p-r) + 2h(p-r-1) \right] \, .
\end{aligned}
\eeq
Similar manipulations can be done for the other possible values attained by the combination $p-r$. 

Combining all the cases and restoring the terms in the second square parenthesis of eq.~\eqref{eq:I4a_starting2}, we find
\beq
\mathcal{I}^{(4)}_{\ref{subfig:1loop_4pt_old_old}}  =
 \frac{i \tilde{g}^4 \, \delta_{p+q,r+s}}{2 N^2 \le \omega_p + \omega_q + p + q + 1 \ri} \,
  \\
   \left( f^{hbv}  f^{eav}  f^{cew}  f^{dhw}  \, 
  \mathcal{P}_{\ref{subfig:1loop_4pt_old_old}}  
  +  f^{hbv}  f^{eav}  f^{chw}  f^{dew}  
   \, \mathcal{Q}_{\ref{subfig:1loop_4pt_old_old}}  \right) \, ,
   \label{eq:I4a_starting3}
\eeq
where 
\begin{subequations}
\beq
   \mathcal{P}_{\ref{subfig:1loop_4pt_old_old}}   =  \left\{
	\begin{aligned}
		&  \frac{1}{p-r}\left[ h(r)-h(p)+h(q)-h(s)+ 2h(p-r) + 2h(p-r-1) \right], & p-r \geq 1\\
        & \frac{1}{r-p}\left[ h(p)-h(r)+h(s)-h(q)+ 2h(r-p) + 2h(r-p-1) \right], & p-r \leq -1 \\
		& \quad \sum_{m=1}^{p} \quad \frac{1}{m^2}+\sum_{m=1}^{q}   \frac{1}{m^2} . & p-r=0
	\end{aligned}
	\right.
    \label{eq:I4a-line2-result}
\eeq
\beq
   \mathcal{Q}_{\ref{subfig:1loop_4pt_old_old}}   =  \left\{
	\begin{aligned}
		&  \frac{1}{p-s}\left[ h(s)-h(p)+h(q)-h(r)+ 2h(p-s) + 2h(p-s-1) \right], & p-s \geq 1\\
        & \frac{1}{s-p}\left[ h(p)-h(s)+h(r)-h(q)+ 2h(s-p) + 2h(s-p-1) \right], & p-s \leq -1 \\
		& \quad \sum_{m=1}^{p} \quad \frac{1}{m^2}+\sum_{m=1}^{q}   \frac{1}{m^2} . & p-s=0
	\end{aligned}
	\right.
    \label{eq:I4a-line2-resul2-t}
\eeq
\end{subequations}

\paragraph{Diagram in fig.~\ref{subfig:1loop_4pt_new_old}.}
The expression~\eqref{eq:term2_4ptvertex_1loop_app} can be treated similarly: we solve the integration over $\omega$ as in eq.~\eqref{eq:integral_4pt_residue}, and then we perform the summations over the integer spatial momenta.
At the end of the computation, we find
\beq
\mathcal{I}^{(4)}_{\ref{subfig:1loop_4pt_new_old}}  =
- \frac{i \tilde{g}^4 \,  \delta_{p+q,r+s}}{2 N^2 \le \omega_p + \omega_q + p + q + 1 \ri} \,
   \left( f^{hbv}  f^{eav}  f^{cew}  f^{dhw}  \, 
  \mathcal{P}_{\ref{subfig:1loop_4pt_new_old}}  
  +  f^{hbv}  f^{eav}  f^{chw}  f^{dew}  
   \, \mathcal{Q}_{\ref{subfig:1loop_4pt_new_old}}  \right) \, ,
   \label{eq:I4b_result}
\eeq
where
\begin{subequations}
\beq
   \mathcal{P}_{\ref{subfig:1loop_4pt_new_old}}   =  \left\{
	\begin{aligned}
		&  \frac{1}{p-r}\left(  h(p) + h(q)   \right), & p-r \geq 1\\
        & \frac{1}{r-p}\left(  h(p) + h(q)   \right), & p-r \leq -1 \\
		& 0 . & p-r=0
	\end{aligned}
	\right.
    \label{eq:I4b-line2-result}
\eeq
\beq
   \mathcal{Q}_{\ref{subfig:1loop_4pt_new_old}}   =  \left\{
	\begin{aligned}
		&  \frac{1}{p-s}\left(  h(p) + h(q)   \right), & p-s \geq 1\\
        & \frac{1}{s-p}\left(  h(p) + h(q)   \right), & p-s \leq -1 \\
		&  0 . & p-s=0
	\end{aligned}
	\right.
    \label{eq:I4b-line2-resul2-t}
\eeq
\end{subequations}

\paragraph{Diagram in fig.~\ref{subfig:1loop_4pt_old_new}.}
We directly report the result:
\beq
\mathcal{I}^{(4)}_{\ref{subfig:1loop_4pt_old_new}}  =
- \frac{i \tilde{g}^4 \, \delta_{p+q,r+s} }{2 N^2 \le \omega_p + \omega_q + p + q + 1 \ri} \,
   \left( f^{hbv}  f^{eav}  f^{cew}  f^{dhw}  \, 
  \mathcal{P}_{\ref{subfig:1loop_4pt_old_new}}  
  +  f^{hbv}  f^{eav}  f^{chw}  f^{dew}  
   \, \mathcal{Q}_{\ref{subfig:1loop_4pt_old_new}}  \right) \, ,
   \label{eq:I4c_result}
\eeq
where
\begin{subequations}
\beq
   \mathcal{P}_{\ref{subfig:1loop_4pt_old_new}}   =  \left\{
	\begin{aligned}
		&  \frac{1}{p-r}\left(  h(r) + h(s)   \right), & p-r \geq 1\\
        & \frac{1}{r-p}\left(  h(r) + h(s)   \right), & p-r \leq -1 \\
		& 0 . & p-r=0
	\end{aligned}
	\right.
    \label{eq:I4c-line2-result}
\eeq
\beq
   \mathcal{Q}_{\ref{subfig:1loop_4pt_old_new}}   =  \left\{
	\begin{aligned}
		&  \frac{1}{p-s}\left(  h(r) + h(s)   \right), & p-s \geq 1\\
        & \frac{1}{s-p}\left(  h(r) + h(s)   \right), & p-s \leq -1 \\
		&  0 . & p-s=0
	\end{aligned}
	\right.
    \label{eq:I4c-line2-resul2-t}
\eeq
\end{subequations}

\paragraph{Diagram in fig.~\ref{subfig:1loop_4pt_new_new}.}
The result reads
\beq
\begin{aligned}
     \mathcal{I}^{(4)}_{\ref{subfig:1loop_4pt_new_new}} & = 
 \frac{i \tilde{g}^4}{2 N^2 \le p+q+\omega_p+\omega_q+1 \ri} \, \le h(p) + h(q) \ri^2 \\
& \times  \left(  f^{hbv} f^{eav} f^{cew} f^{dhw}  
 \delta_{p,r} \delta_{q,s}   
 - f^{hbv} f^{eav} f^{chw} f^{edw} \delta_{p,s} \delta_{q,r} \right) \, .
\end{aligned}
\label{eq:result_I4d}
\eeq
Summing the terms in eqs.~\eqref{eq:I4a_starting3}, \eqref{eq:I4b_result}, \eqref{eq:I4c_result} and \eqref{eq:result_I4d}, we obtain the full one-loop correction reported in eq.~\eqref{eq:I4_result}.

\subsection{Details on the QM computation of $2 \rightarrow 2$ scattering amplitude}
\label{app:ssec:QM_computation_22}

In this section, we show that performing the summations $U^{ij}_{rs} U^{ij}_{pq}$ and $U^{ij}_{sr} U^{ij}_{pq}$ in eq.~\eqref{eq:step1_1loop_quartic_QM} leads to the expression in eq.~\eqref{eq:step2_1loop_quartic_QM}.
To this aim, we need to consider three possible subcases depending on the momenta of the external particles:

\paragraph{Case 1 ($p=r$).}
The total momentum conservation $p+q=r+s$ implies $q=s$ as well.
Using the dictionary~\eqref{eq:dictionary_U_su11}, we find by direct computation
\beq
    U^{ij}_{pq} U^{ij}_{pq} =\frac{1}{4} \left[ (h(p)+h(q))^2+\sum_{l=1}^{p} \left(\frac{1}{l} \right)^2+\sum_{l=1}^{q} \left(\frac{1}{l} \right)^2 \right] \, .
\eeq
This result matches with the last line in eq.~\eqref{eq:I4-line2-result}.

\paragraph{Case 2 ($p-r \geq 1$).}
Let us parametrize $p=r+\Delta$, where $\Delta \geq 1$. Using the dictionary~\eqref{eq:dictionary_U_su11}, we find
\beq
\begin{aligned}
 U^{ij}_{rs} U^{ij}_{pq} & = - \frac{1}{2} U^{pq}_{rs} (h(p)+h(q))  + \sum_{l=1}^{p} \frac{1}{2l}  U^{p-l,q+l}_{rs}  + \sum_{l=1}^{q} \frac{1}{2l} U^{p+l,q-l}_{rs}  = \\
& =\frac{1}{2} \left[ -\frac{1}{2 \Delta} (h(p)+h(q)) + \sum_{l=1}^{\Delta-1} \frac{1}{2 (\Delta-l)} \frac{1}{l} -(h(r)+h(s))\frac{1}{2\Delta} \right. \\
& \left. + \sum_{l=\Delta+1}^{p} \frac{1}{2 (l-\Delta)} \frac{1}{l} + \sum_{l=1}^{q} \frac{1}{2 (\Delta+l)} \frac{1}{l}  \right]  = \\
&= \frac{1}{4 \Delta} \left[ -h(p)-h(q)+ h(\Delta-1) +h(\Delta-1)-h(r)-h(s) \right.\\
& \left.+h(p-\Delta)-h(p)+h(\Delta)+h(q)-h(q+\Delta)+h(\Delta) \right] =\\
& = \frac{1}{4 (p-r)} \left[ -2 h(p)-2 h(s) +2 h(p-r)+ 2 h(p-r-1)\right] \, .
\end{aligned}
\eeq
This expression matches with the first line in eq.~\eqref{eq:I4-line2-result}.

\paragraph{Case 3 ($p-r \leq 1$).}
With similar manipulations, we get:
\beq
\begin{aligned}
U^{ij}_{rs} U^{ij}_{pq} & = - \frac{1}{2} U^{pq}_{rs} (h(p)+h(q)) + \sum_{l=1}^{p} \frac{1}{2l} U^{p-l,q+l}_{rs}  + \sum_{l=1}^{q} \frac{1}{2l} U^{p+l,q-l}_{rs} = \\
& = \frac{1}{4 \Delta} \left[-2h(r)-2h(q)+2h(r-p)+2h(r-p-1)\right] \, ,
\end{aligned}
\eeq
which matches with the last line in eq.~\eqref{eq:I4-line2-result}.

By exchanging $r \leftrightarrow s$, we can similarly reproduce the contribution~\eqref{eq:I4-line2-resul2-t}.

\addcontentsline{toc}{section}{References}

\bibliography{SU11SMT}
\bibliographystyle{JHEP}
\end{document}